\documentclass{aa}
\usepackage{graphicx}
\usepackage{natbib}
\usepackage{txfonts}
\input epsf

\begin{document}

   \title{Challenges for asteroseismic analysis of Sun-like stars}

   \author{W.~J.~Chaplin\inst{1}
          \and
          G.~Houdek\inst{2}
	  \and
          T.~Appourchaux\inst{3}
          \and
          Y.~Elsworth\inst{1}
	  \and
          R.~New\inst{4}
	  \and
          T.~Toutain\inst{1}
	  }

   \offprints{W.~J.~Chaplin}

   \titlerunning{Asteroseismic analysis of Sun-like stars}

   \authorrunning{Chaplin et al.}

   \institute{School of Physics and Astronomy,
              University of Birmingham, Edgbaston,
              Birmingham B15 2TT, U.K\\
              \email{w.j.chaplin@bham.ac.uk}
         \and
              Institute of Astronomy,
              University of Cambridge,
              Cambridge CB3 0HA, UK
              \email{hg@ast.cam.ac.uk}
         \and
             Institut d'Astrophysique Spatiale (IAS),
             Batiment 121, F-91405, Orsa Cedex, France
         \and 
              Faculty of Arts, Computing,
              Engineering and Sciences, Sheffield Hallam
              University, Sheffield S1 1WB
             }

  \date{...}

  \abstract
  {Asteroseismology of Sun-like stars is undergoing rapid expansion
  with, for example, new data from the CoRoT mission and continuation
  of ground-based campaigns. There is also the exciting upcoming
  prospect of NASA's Kepler mission, which will allow the
  asteroseismic study of several hundred Sun-like targets, in some
  cases for periods lasting up to a few years.}
  {The seismic mode parameters are the input data needed for making
  inference on stars and their internal structures. In this paper we
  discuss the ease with which it will be possible to extract estimates
  of individual mode parameters, dependent on the mass, age, and
  visual brightness of the star. Our results are generally applicable;
  however, we look at mode detectability in the context of the
  upcoming Kepler observations.}
  {To inform our discussions we make predictions of various seismic
  parameters. To do this we use simple empirical scaling relations and
  detailed pulsation computations of the stochastic excitation and
  damping characteristics of the Sun-like p modes.}
  {The issues related to parameter extraction on individual p modes
  discussed here are mode detectability, the detectability and impact
  of stellar activity cycles, and the ability to measure properties of
  rotationally split components, which is dependent on the relative
  importance of the rotational characteristics of the star and the
  damping of the stochastically excited p modes.}
  {}


 \maketitle

 \section{Introduction}
 \label{sec:intro}

With the recent launch of CoRoT (Baglin et al. 2006), the upcoming
launch of NASA's Kepler mission (Basri et al. 2005) and continuation
of observations by MOST (Matthews et al. 2007) and ground-based teams
(Bedding \& Kjeldsen 2006), we are entering a rewarding era for
asteroseismic studies of the interiors of Sun-like stars. Sun-like
oscillations give a very rich spectrum allowing internal structures
and dynamics to be probed down into the stellar cores to very high
precision. Asteroseismic observations of many stars will allow
multiple-point tests of crucial aspects of stellar evolution and
dynamo theory.

The prospects for asteroseismology with Kepler are particularly
exciting. In addition to searching for Earth-like exoplanets (via the
transit method), the Kepler Asteroseismology Investigation (KAI) --
which is arranged around the Kepler Asteroseismic Science Operations
Centre (KASOC) -- will provide an unprecedented opportunity to study
several hundred stars showing Sun-like oscillations
(Christensen-Dalsgaard et al. 2007, 2008). Because the nominal mission
lifetime is 3.5\,years we should expect to have long datasets on many
of these stars.

The input data for probing stellar interiors are the mode parameters,
such as individual frequencies, frequency splittings, linewidths, and
powers. For those stars where measurement of individual mode
parameters is difficult, the input data will be ensemble averages,
e.g., mean frequency spacings. Accurate mode parameter data are a
vital prerequisite for robust, accurate inference on the internal
structures of the stars. Asteroseismic observations covering periods
of several years will also allow studies of stellar cycles from
observation of systematic stellar-cycle-driven variations of the mode
parameters, thereby giving important information to the dynamo
theorists.

Parameters of individual modes may be estimated using the
``peak-bagging'' fitting techniques, which have been applied to
Sun-as-a-star helioseismology data (e.g., Chaplin et al. 2006) and
which are now being applied to Sun-like asteroseismology data, e.g.,
the CoRoT data (Appourchaux et al. 2006a, b) and also WIRE spacecraft
data (Fletcher et al. 2006; Karoff et al. 2007). Peak-bagging involves
maximum-likelihood fitting of mode peaks in the frequency power
spectrum to multi-parameter fitting models, where individual mode
peaks are represented by Lorentzian-like functions. The parameters may
also be extracted using methods which have been developed for analysis
of ground-based data on Sun-like stars (e.g., see Bedding et al. 2004;
Kjeldsen et al. 2005). For example, from measurement of the variance,
over time, of frequency locations of peaks it is possible to estimate
the intrinsic damping rates of stochastically excited modes.

In this paper we are concerned with making predictions of the ease
with which it will be possible to extract estimates of parameters on
individual low-degree (low-$l$) acoustic (p) modes, in different
regions of the HR diagram occupied by the Sun-like oscillators. We do
so with an eye on the upcoming Kepler observations.  To make our
predictions we use simple empirical scaling relations, together with
detailed pulsation computations of the stochastic excitation and
damping characteristics of the p modes. The predictions and
discussions that follow are by no means exhaustive. However, they do
cover several important issues that are relevant to the challenge of
extracting estimates of parameters on individual modes. The main
issues we consider, for stars of different mass and age, are:
\begin{itemize}

 \item The detectability of modes;

 \item The impact of stellar activity cycles on the observed mode
 peaks. We comment on detectability of those cycles; we also consider
 changes to the mode power and damping rates, which have implications
 not only for detectability but also detailed comparison of
 observations with theoretical predictions of the excitation and
 damping; and

 \item The ability to resolve individual components in the non-radial
 mode multiplets, which depends on the relative importance of rotation
 and mode damping.

\end{itemize}

The layout of our paper is as follows. We begin in
Section~\ref{sec:data} with a brief description of the pulsation
calculations, which are performed for a grid of 31 stellar models.  We
then look in Section~\ref{sec:sn} at predictions of the mode powers
and widths for the models and discuss detectability issues in the
context of noise levels expected for the Kepler observations, which
will be photometric. In Section~\ref{sec:cyc}, we use empirical
scaling relations to predict stellar-cycle characteristics for our
grid of stellar models and use these predictions to comment on the
variability we might expect in estimates of p-mode frequencies, powers
and damping rates.  Our discussion on stellar-cycle variability also
includes a look at how shapes of mode peaks in the frequency power
spectrum may be distorted by strong cycles. Finally, in
Section~\ref{sec:rot} we use empirical predictions of surface rotation
rates to comment on the difficulty of resolving components in
non-radial mode multiplets.  We then draw together the main summary
points in Section~\ref{sec:sum}.

 \section{Stellar model data}
 \label{sec:data}

\begin{figure*}
 \centerline {\epsfxsize=8.0cm\epsfbox{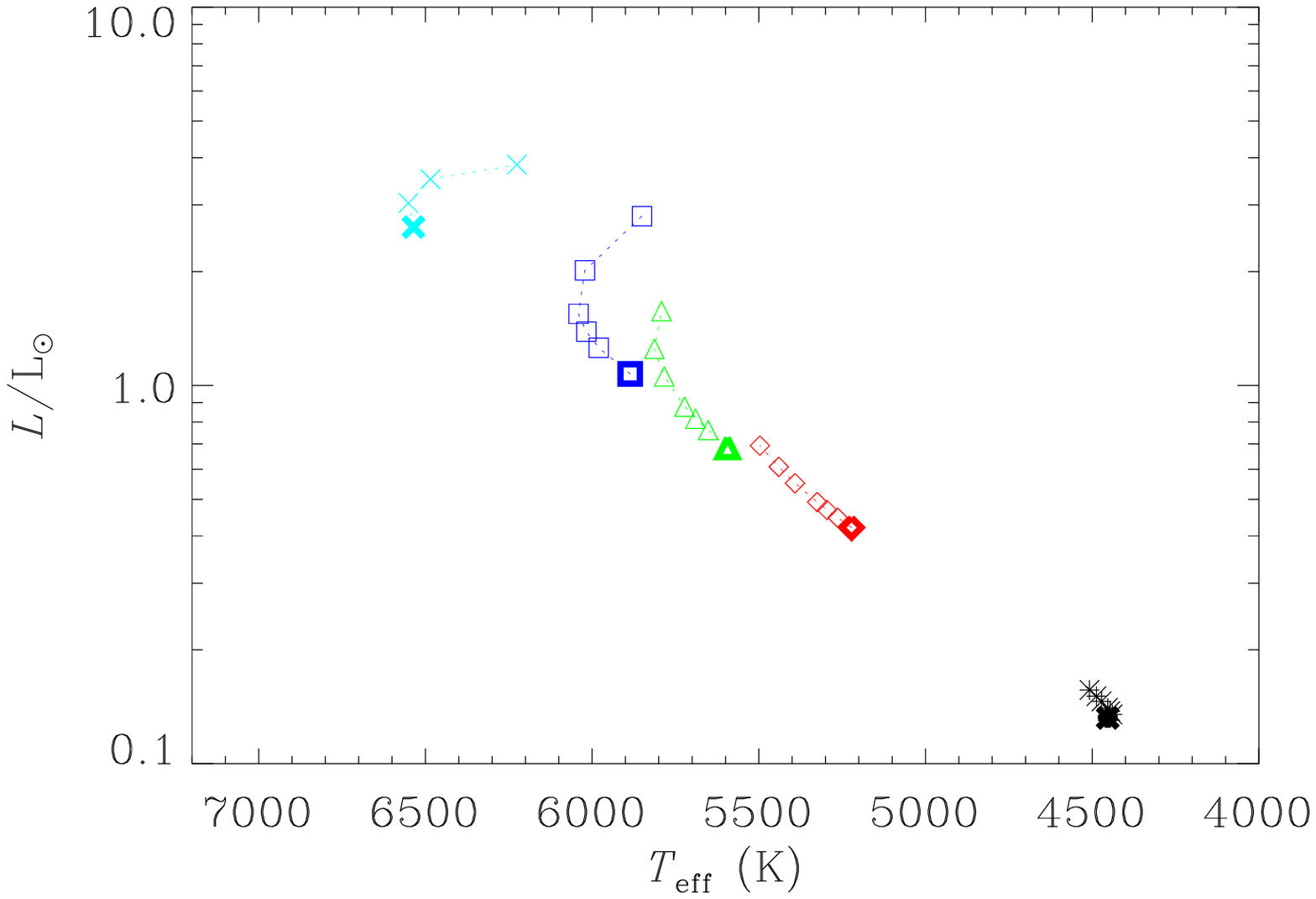}
              \epsfxsize=8.0cm\epsfbox{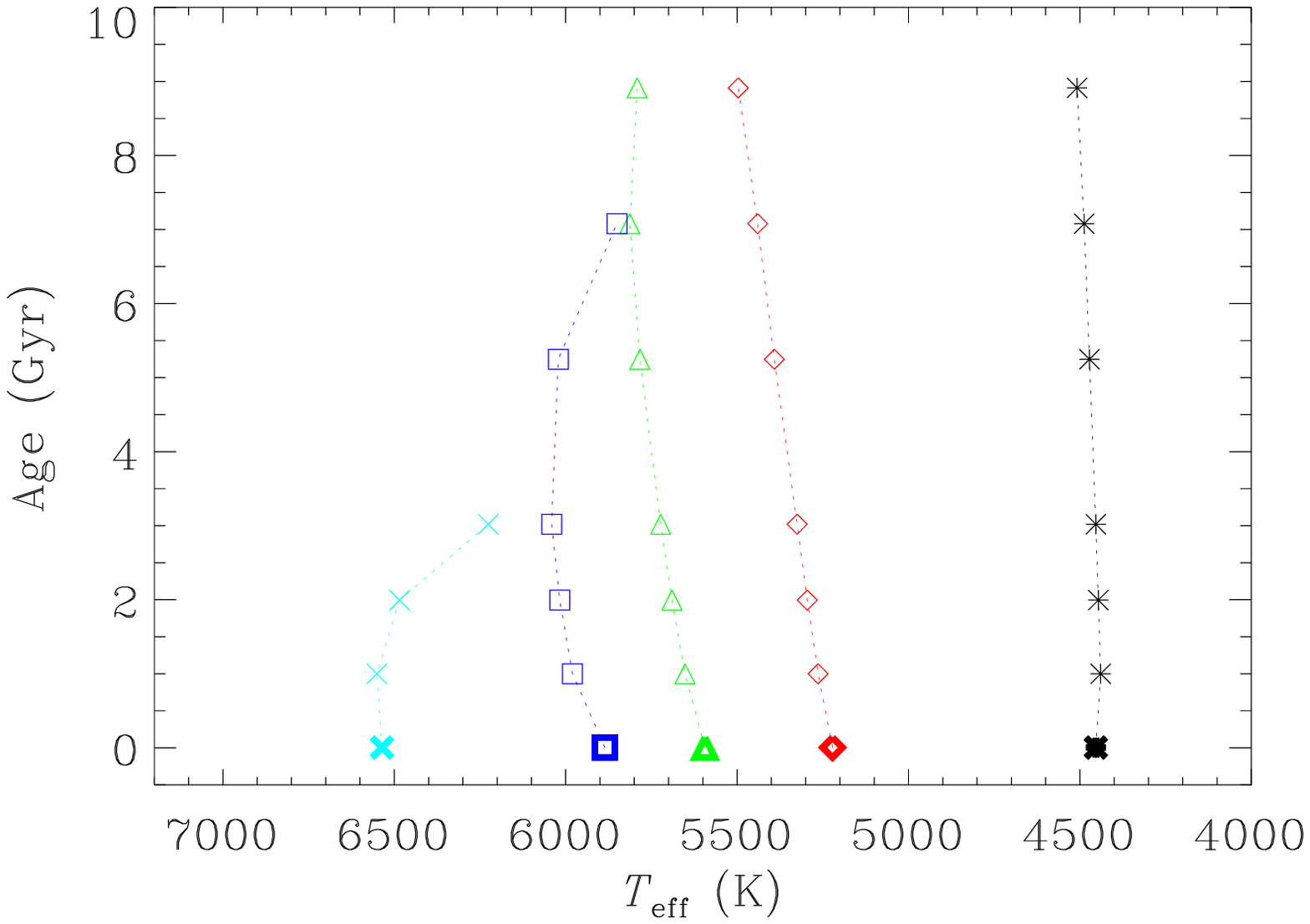}}

 \caption{Luminosity-effective temperature plot (left-hand panel) and
age-effective temperature plot (right-hand panel) for a sequence of
stellar models. Different symbols denote models of different mass:
$0.7\,\rm M_{\odot}$ as black stars; $0.9\,\rm M_{\odot}$ as red
diamonds; $1.0\,\rm M_{\odot}$ as green triangles; $1.1\,\rm
M_{\odot}$ as blue squares; $1.3\,\rm M_{\odot}$ as diagonal cyan
crosses. Evolutionary sequences are joined with a dotted line. The
same symbol and colour scheme is used in other plots in the
paper. Data on the ZAMS models are rendered with bold symbols.}

 \label{fig:HR}
\end{figure*}

We have considered a grid of stellar models, with masses in the range
0.7 to $1.3\,\rm M_{\odot}$ and ages in the range from the ZAMS to
9\,Gyr.  We used the Padova isochrones (Bonatto, Bica \& Girardi 2004;
Girardi et al. 2002, 2004) to specify the primary characteristics of
each model, i.e., mass $M$, radius $R$, effective temperature $T_{\rm
  eff}$, and luminosity $L$. The composition was fixed at $X=0.7$ and
$Z=0.019$ for all models. Fig.~\ref{fig:HR} shows a
luminosity-effective temperature plot (left-hand panel) and an
age-effective temperature plot (right-hand panel), for the models. The
different symbols denote models of different mass (see caption).

The stellar equilibrium and pulsation computations that we performed
are as described by Balmforth (1992), Houdek et. al (1999), and
Chaplin et al. (2005).  The pulsation computations required estimates
of $M$, $T_{\rm eff}$, $L$ and the composition as input. The
computations gave as output estimates of the acoustic powers and
damping rates of the radial p modes of each stellar model. We
essentially followed the same recipe to perform the computations as
Chaplin et al. (2007). However, whereas in Chaplin et al. we performed
computations for 22 stellar models whose input parameters were chosen
to match those of stars that have been observed by the Mount Wilson
Ca~II H\&K program, here we instead performed computations for the
grid of 31 models having basic parameters to reflect systematic
variations in mass and age.


 \section{Mode heights and widths in the frequency power spectrum}
 \label{sec:sn}

Information on the p modes may be extracted by, for example, analysis
of the p-mode peaks in the frequency power spectrum. Here, we use the
results of the stellar model computations to make predictions of the
peak parameters. The stellar model computations provided predictions
of two independent sets of radial-mode parameters for each star: the
linear damping rates, $\eta$, and the excitation rates (acoustic
powers), $P$. The observed parameters of the mode peaks in the
frequency power spectrum are formed from these quantities. The peak
\textsc{fwhm} linewidths are given by $\Gamma = \eta / \pi$ while the
velocity powers, $V^2$, are calculated from (e.g., Houdek et
al. 1999):
 \begin{equation}
 V^2 = \frac{P}{2\eta I}.
 \label{eq:V2}
 \end{equation}
Here, $I$ is the mode inertia. The computations were calibrated so
that for a model of the Sun the maximum \textsc{rms} mode amplitude was
$V=0.16\,\rm ms^{-1}$.

In order to predict mode amplitudes, $A$, in intensity -- to allow us
to discuss results in the context of the photometric Kepler
observations -- we convert from the velocity amplitudes, $V$, using
(Kjeldsen \& Bedding 1995):
 \begin{equation}
 A = (dL/L)_{\lambda} = \frac{V\,/\rm m\,s^{-1}}{(\lambda/550\,\rm
 nm)(T_{\rm eff}/5777\,\rm K)^2}\,20.1\,\rm ppm.
 \label{eq:a}
 \end{equation}
The spectral \textsc{fwhm} bandpass of Kepler runs from 430 to
890\,nm. We therefore used an average wavelength of $\lambda =
660\,\rm nm$ in Equation~\ref{eq:a} above.

If observations are of sufficient length to resolve mode peaks in the
frequency power spectrum the $A^2$ (or $V^2$) do not give the observed
\emph{heights} of the peaks. The requirement for peaks to be resolved
may be put in terms of the mode lifetime $\tau = 1 / \eta$. The
requirement is that the observation length $T \gg 2\tau$ (Chaplin et
al. 2003). When the peaks are resolved, the heights -- or
\emph{maximum power spectral densities} -- are instead given by
(Chaplin et al. 2005):
 \begin{equation}
 H = \frac{2A^2}{\pi \Gamma} = \frac{P}{\eta^2 I}.
 \label{eq:height}
 \end{equation}
In what follows as shall assume the condition $T \gg 2\tau$ is met for
the observations. We make some comments on this assumption in
Appendix~\ref{sec:hall}. Furthermore, we use \textsc{rms} amplitudes,
$A$, to calculate the heights $H$.

 \subsection{Frequency at maximum mode height}
 \label{sec:maxh}


\begin{figure}
 \centerline {\epsfxsize=8.0cm\epsfbox{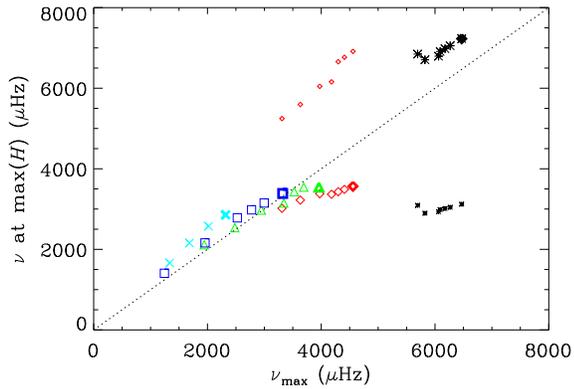}}

 \caption{Frequency $\nu$ at which the mode height, $H$, of the models
is a maximum, plotted against the frequency $\nu_{\rm max}$ resulting
from the scaling in Equation~\ref{eq:maxnu}. The dotted line shows the
locus $\nu = \nu_{\rm max}$. Data on the ZAMS models are rendered with
bold symbols.}

 \label{fig:maxnu}
\end{figure}


It has been shown that for the Sun-like oscillators, the frequency of
maximum power of the p modes scales to good approximation with the
acoustic cut-off frequency (Bedding \& Kjeldsen 2003). The frequency
of maximum power for the radial solar p modes is $\sim 3100\,\rm \mu
Hz$. We use this value to calibrate the following scaling relation:
 \begin{equation}
 \nu_{\rm max} = \frac{M/\rm M_{\odot}}{(R/\rm R_{\odot})^2
 \sqrt{T_{\rm eff}/5777\,\rm K}}\,3100\,\rm \mu Hz.
 \label{eq:maxnu}
 \end{equation}
How well do the pulsation computations match this relation? We
consider the match for computations of the heights, $H$, as opposed to
the amplitudes, $A$, since it is the heights that determine the
detectability of modes in the frequency power spectrum.

\begin{figure*}
 \centerline {\epsfxsize=8.0cm\epsfbox{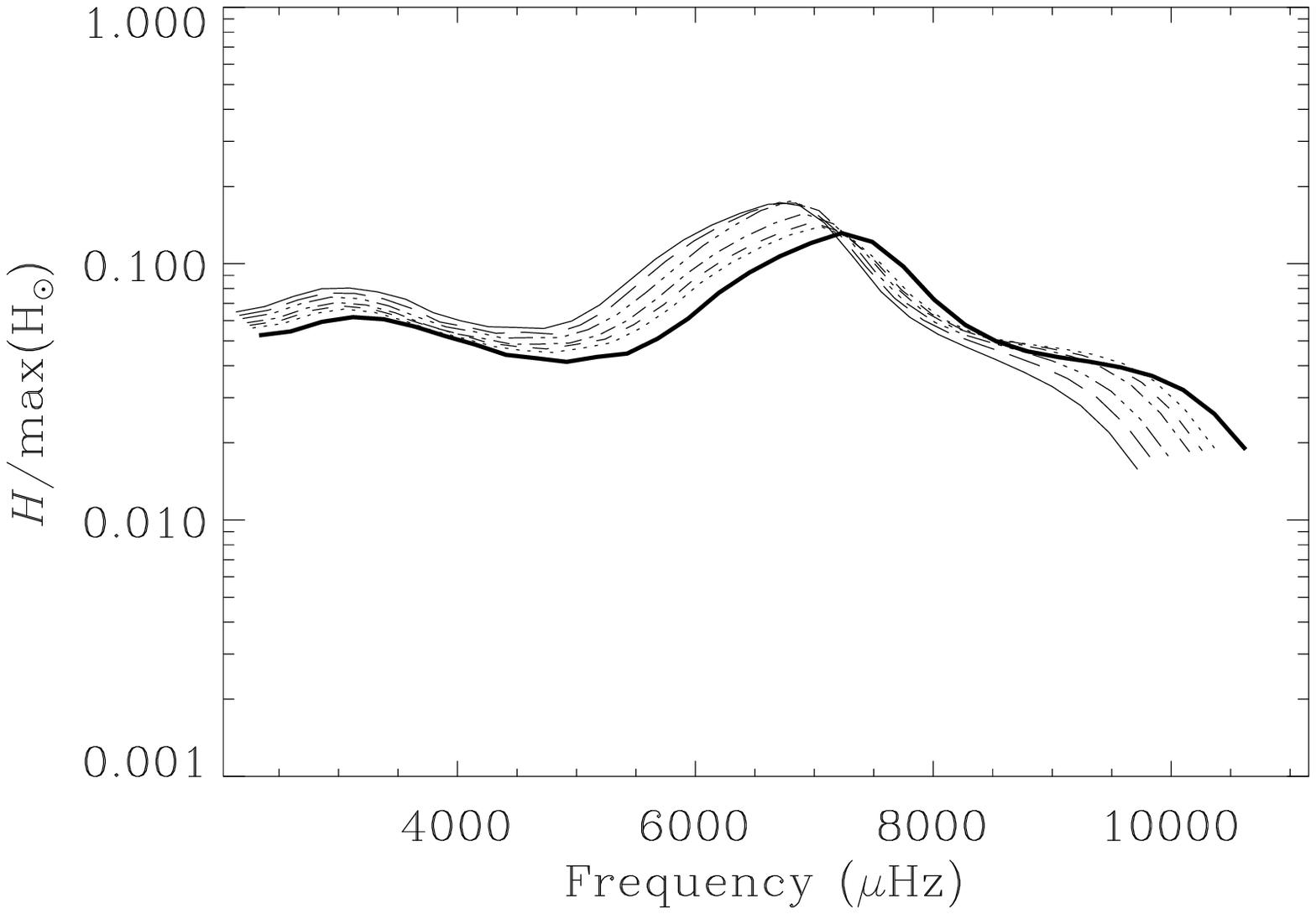}
              \epsfxsize=8.0cm\epsfbox{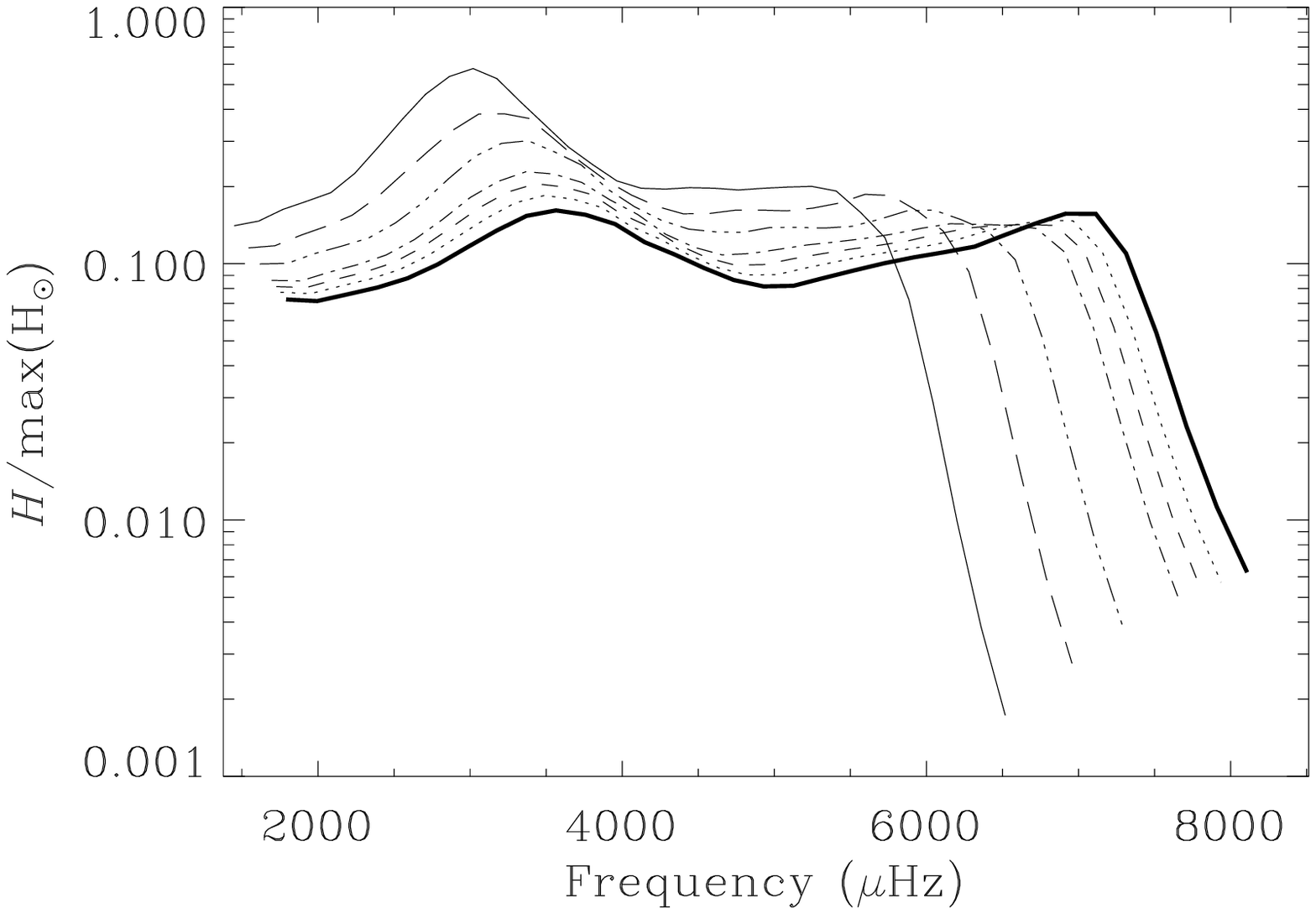}}

 \caption{Predicted height envelope of the radial modes for the
 $M=0.7\,\rm M_{\odot}$ models (left-hand panel) and the $M=0.9\,\rm
 M_{\odot}$ models (right-hand panel) versus frequency (see
 text). Results on the ZAMS models are rendered with a thick solid
 line, while the results on the oldest 9-Gyr models are rendered with
 a thin solid line.}

 \label{fig:hall}
\end{figure*}

In Fig.~\ref{fig:maxnu} we plot the frequency $\nu$ of maximum height
$H$ from the model computations against the scaled frequency $\nu_{\rm
max}$ from Equation~\ref{eq:maxnu}. The dotted line shows the locus
$\nu = \nu_{\rm max}$ along which the pulsation computation results
would lie should they follow Equation~\ref{eq:maxnu}. While the match
is reasonable for most of the models, there are results for two
sequences of models which lie a long way from the dotted line. These
are results on the lighter $0.7\,\rm M_{\odot}$ and $0.9\,\rm
M_{\odot}$ models, where the pulsation computations actually show
\emph{two} maxima in $H$ (see Fig.~\ref{fig:hall}). We plot data in
Fig.~\ref{fig:maxnu} for the two maxima at each mass: at both masses
one maximum lies close to the $\nu = \nu_{\rm max}$ line, while the
other does not.

The presence of a second maximum is due predominantly to local
depressions in the linear damping rates (as a function of frequency),
as was pointed out by Chaplin et al. (2007). For the $0.7\,\rm
M_{\odot}$ models, $H$ at the higher-frequency maximum is always
larger than $H$ at the lower-frequency maximum (see left-hand panel of
Fig.~\ref{fig:hall}). The frequency locations of these
higher-frequency maxima are plotted as full-sized asterisks in
Fig.~\ref{fig:maxnu} and are seen to lie closer to the $\nu = \nu_{\rm
max}$ curve than their out-of-line, lower-frequency counterparts (the
latter plotted as small asterisks). For the $0.9\,\rm M_{\odot}$
models it is the lower-frequency maxima that are the more prominent
(see right-hand panel of Fig.~\ref{fig:hall}), and which also follow
more closely the linear scaling with $\nu_{\rm max}$ (see full-sized
diamonds in Fig.~\ref{fig:maxnu}). In what follows we use pulsation
computation results on modes at the strongest maxima.


 \subsection{Height-to-background ratios in Kepler data}
 \label{sec:kepler}

In order to give some observational context for the computed heights,
$H$, here we compare them with shot noise power spectral densities
predicted for intensity observations by the Kepler instrument.

\begin{figure*}
 \centerline {\epsfxsize=8.0cm\epsfbox{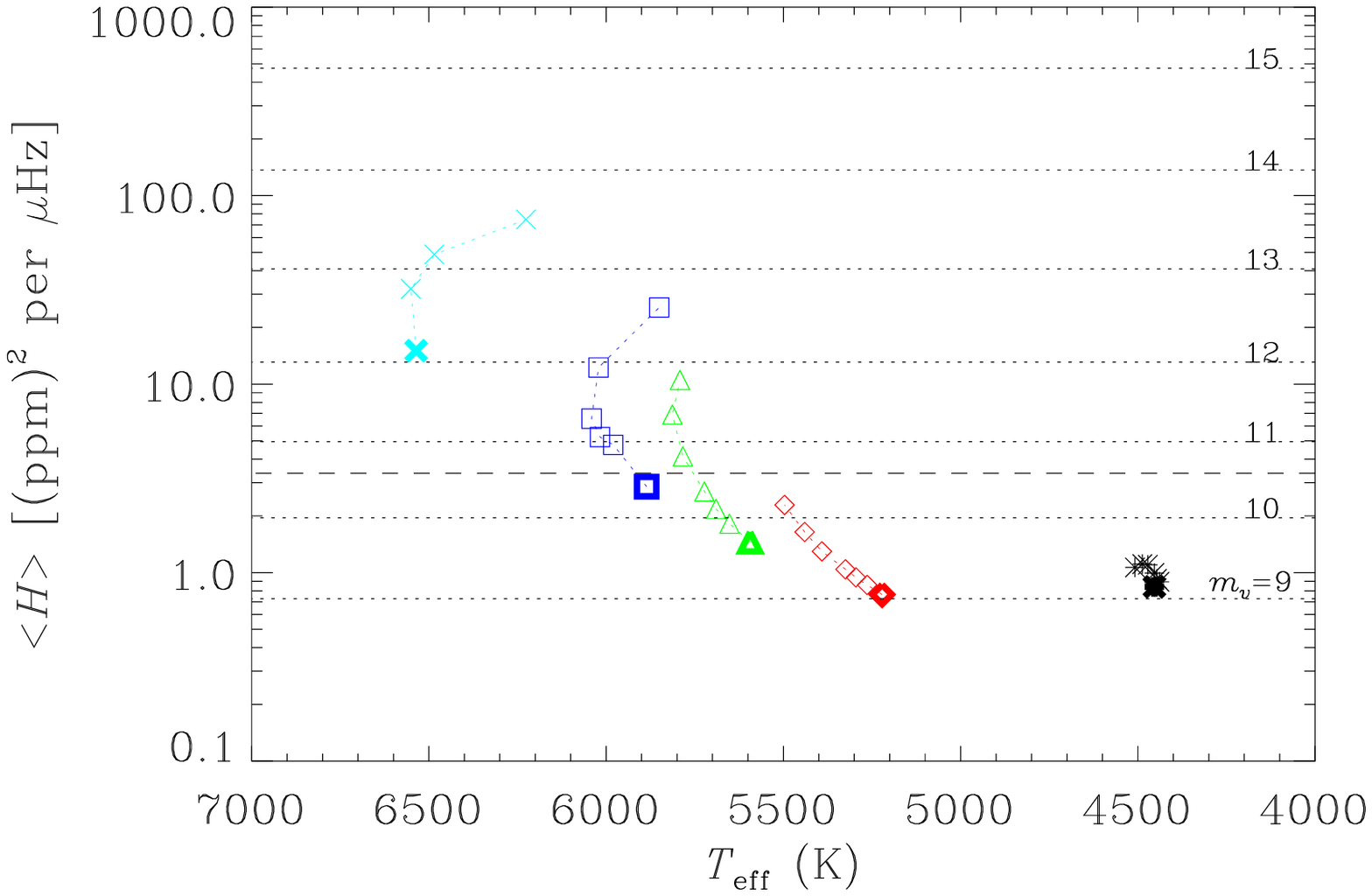}
              \epsfxsize=8.0cm\epsfbox{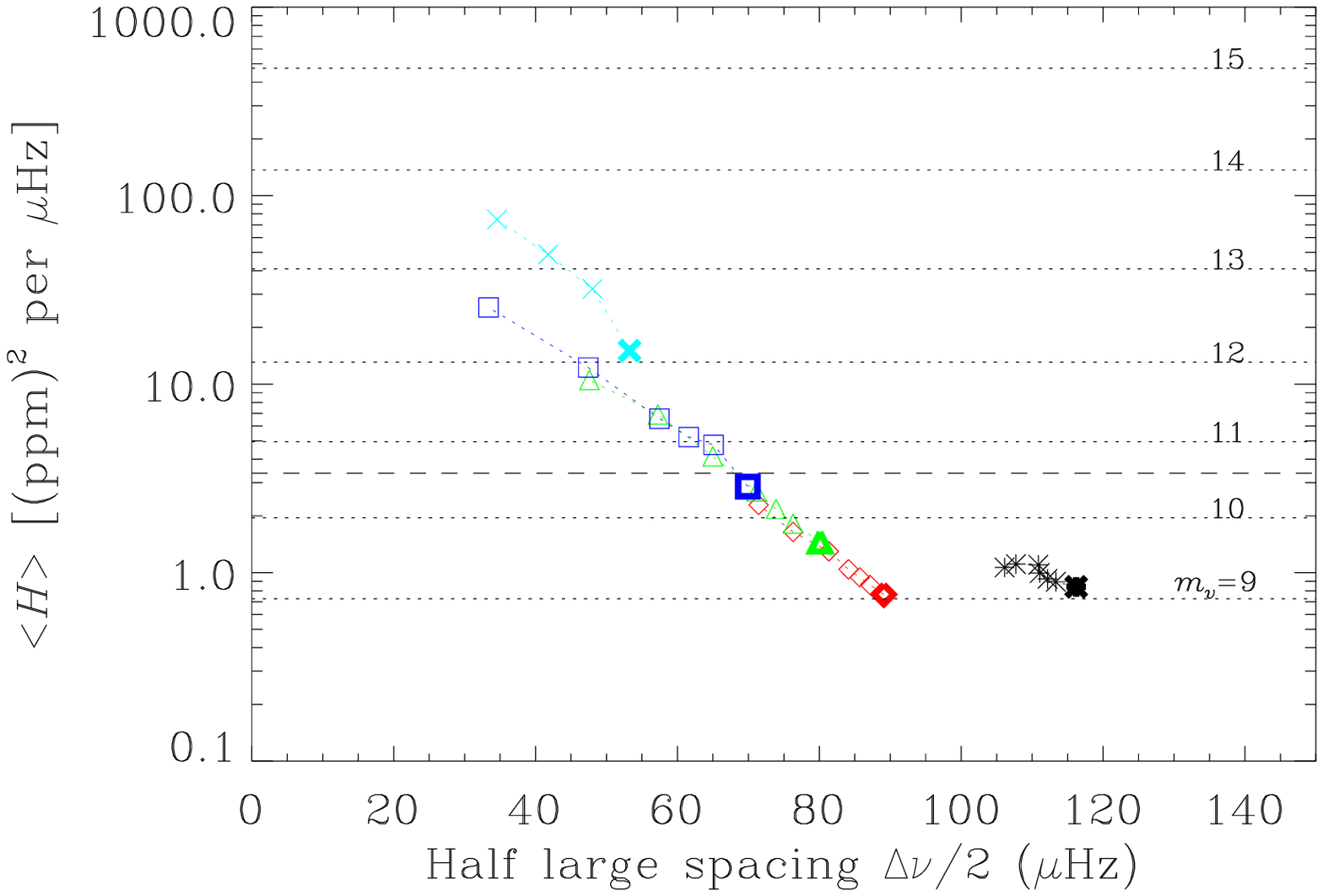}}

 \caption{Average maximum power spectral density, or height, $\left< H
 \right>$ of the five strongest modes (same symbols as
 Fig.~\ref{fig:HR}), plotted as a function of effective temperature
 (left-hand panel) and half the average large frequency spacing
 (right-hand panel). The dashed lines show the average for the five
 strongest solar radial modes. The dotted lines are expected shot
 noise power spectral densities for Kepler, for $m_v=9$ (lowest-lying
 line) to 15 (top line). Data on the ZAMS models are rendered with
 bold symbols.}

 \label{fig:noi}
\end{figure*}

First, we follow the procedure in Chaplin et al. (2007) and average
calculated heights from the pulsation computations over the five
strongest radial modes, to give $\left< H \right>$. Fig.~\ref{fig:noi}
shows the resulting estimates of $\left< H \right>$, in units of $\rm
(ppm)^2\,\mu Hz^{-1}$.  The dashed line shows the average for the five
strongest solar radial modes. By averaging results we get predictions
for the detectability of several overtones. Also plotted (dotted
lines) are the noise power spectral density levels, $B$, expected for
Kepler observations at $m_v=9$ (lowest-lying line) to 15 (top
line). These levels were calculated from:
 \begin{equation}
 B = 2 \sigma^2 \Delta t, 
 \label{eq:n}
 \end{equation}
where $\Delta t$ is the cadence (60\,sec) and $\sigma$ are the
estimated Kepler shot noise levels, per 60-sec sample (Kjeldsen,
private communication). The selected $m_v$ span the nominal apparent
magnitude target range for Kepler.

We have plotted $\left< H \right>$ and $B$ against not only effective
temperature $T_{\rm eff}$ (left-hand panel of Fig.~\ref{fig:noi}), but
also half the average large frequency spacing $\Delta \nu / 2$
(right-hand panel). This is the dominant frequency spacing of the
low-$l$ frequency power spectrum, it being approximately the spacing
between consecutive odd-$l$ ($l=1$ and 3) and even-$l$ ($l=0$ and 2)
mode pairs.

\begin{figure*}
 \centerline {\epsfxsize=8.0cm\epsfbox{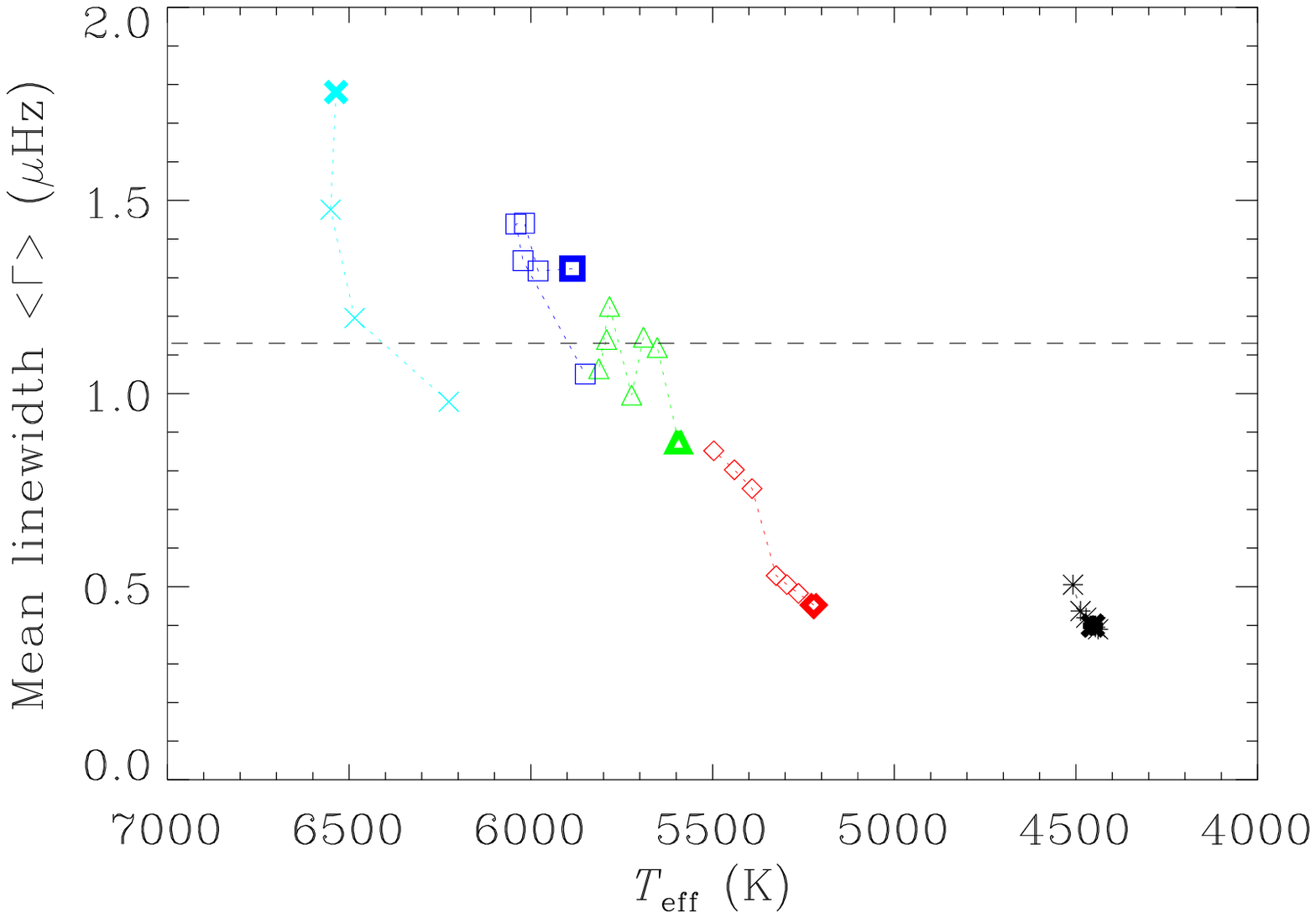}
              \epsfxsize=8.0cm\epsfbox{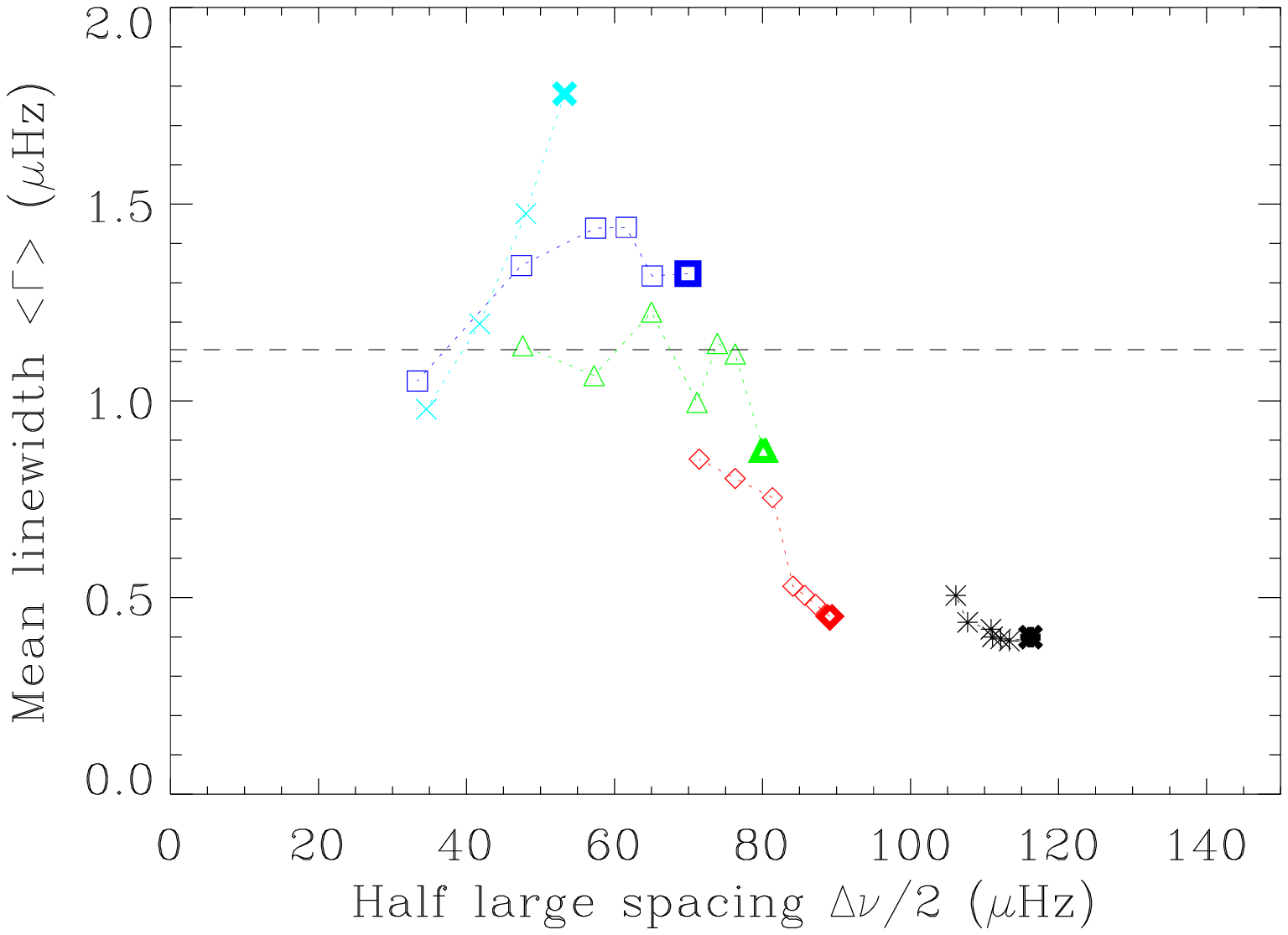}}

 \caption{Average linewidth, $\left< \Gamma \right>$, of the five
 strongest modes (symbols), plotted as a function of effective
 temperature (left-hand panel) and half the average large frequency
 spacing (right-hand panel). The dashed lines show the average for the
 five strongest solar radial modes. Data on the ZAMS models are
 rendered with bold symbols.}

 \label{fig:wid}
\end{figure*}

Fig.~\ref{fig:wid} shows similar plots, this time for the
average linewidths, $\left< \Gamma \right>$, from the pulsation
computations. These averages were again made over the five strongest
modes in each model.


Let us consider some of the important points to take from
Figs.~\ref{fig:noi} and~\ref{fig:wid}:
\begin{itemize}

 \item While the average heights, $\left< H \right>$, increase by
 approximately two orders of magnitude going from the $0.7\,\rm
 M_{\odot}$ to the $1.3\,\rm M_{\odot}$ models, the average linewidths
 $\left< \Gamma \right>$, in contrast increase by only approximately a
 factor of three to four. For a given evolutionary sequence average
 heights increase with increasing age. The behaviour of the average
 linewidths, $\left< \Gamma \right>$, is more complicated. Up to mass
 $1.0\,\rm M_{\odot}$, each evolutionary sequence shows an overall
 increase in average linewidth with increasing age. The changes can be
 far from monotonic and are typically quite modest, e.g., for the
 $1.0\,\rm M_{\odot}$ sequence the change in $\left< \Gamma \right>$
 from the ZAMS to 9\,Gyr amounts to only about 20\,\%. At higher mass
 things get more complicated. The average linewidths of the $1.1\,\rm
 M_{\odot}$ sequence first increase, then decrease. At $1.3\,\rm
 M_{\odot}$ there is a pronounced decrease in average linewidth with
 increasing age.

 \item The best potential targets for extracting estimates of
 individual mode parameters are of course the more massive models, for
 which we expect to have good height-to-background ratios in the
 frequency power spectrum. However, this comes at a cost. First, the
 tendency is for the damping to be stronger in these more massive
 models, compared to their lighter counterparts. Since mode peaks are
 therefore wider in the frequency power spectrum, the data will be
 more susceptible to blending effects between adjacent mode
 components, making clean extraction of frequency spacings or
 splittings harder. The problem of blending can cause severe problems
 for attempts to estimate the angle of inclination offered by the
 star. The inclination affects the relative amplitudes of rotationally
 split components in non-radial modes, meaning it can in principle be
 constrained if the amplitude ratios can be measured. This will not be
 possible in the presence of significant blending. Convergence to, or
 use of, poor estimates of the inclination can also significantly bias
 estimates of the frequency splittings. Even a factor-of-two
 difference in the widths can have a big impact on the analysis (e.g.,
 see Gizon \& Solanki 2003; Ballot et al. 2006, 2008).

 While the magnitudes of the linewidths do give some guide to
 potential problems, a much more useful measure is the ratio of the
 rotational frequency splittings to the linewidths. This ratio
 determines how easy it is to resolve adjacent components in the
 non-radial modes. We go on to discuss issues concerning rotation,
 including predictions of the splitting-to-linewidth ratio, later in
 Section~\ref{sec:rot}.

 The second problem for the more massive models is that they show the
 smallest large frequency spacings. This means power from adjacent
 outlying overtones can have a significant contribution in the
 frequency neighbourhood of a mode whose properties are being
 estimated. A common peak-bagging approach is to go through the
 frequency power spectrum fitting a mode pair at a time. This is
 because the $l=0$ modes lie in close proximity in frequency to the
 $l=2$ modes. The same is true for the $l=1$ and $l=3$ modes. Since
 the fitting models usually only include power from the target pair,
 power in the target frequency window from other outlying modes --
 which is not accounted for in the fitting models -- can bias the
 best-fitting parameters (e.g., see discussion in Fletcher et
 al. 2008). A way round this is of course to take account of the
 outlying power (e.g., Gelly et al. 2002; Jimen\'ez, Roca-Cort\'es \&
 Jimen\'ez-Reyes 2002; Fletcher et al. 2008), or to fit all modes in
 the frequency power spectrum in one go (e.g., see Lazrek et al.,
 2001; Appourchaux 2003; Jefferies, Vorontsov \& Giebink
 2004). Techniques like these may need to be called into play when
 long Kepler datasets are analyzed on the more massive Sun-like
 oscillators.

 \item From ground-based observations of the p-mode spectrum of
 Procyon is has been established that analytical predictions of the
 mode amplitudes \emph{overestimate} the observed amplitudes in this
 hottest part of the `Sun-like' regime, e.g., see the discussion in
 Houdek (2006) and references therein. The discrepancy is
 approximately a factor of five in power for Procyon, which has an
 effective temperature of $T_{\rm eff} \sim 6530\,\rm K$ (Allende
 Prieto et al. 2002). Even if one were to reduce the heights of the
 $1.3\,\rm M_{\odot}$ models (diagonal crosses) in this paper by a
 factor of five the expected height-to-background ratios would remain
 healthy for observations at the bright end of the $m_v$ target range
 of Kepler. Some of the discrepancy between the pulsation computations
 and the observations is most likely due to an underestimation of the
 damping rates by the computations. As such, the mode widths may if
 anything be larger in this part of the HR diagram than the
 predictions shown in Fig.~\ref{fig:wid}, which may add to
 complications for parameter estimation of individual modes.

 \item The height and noise data in Fig.~\ref{fig:noi} suggest it may
 not be possible to extract parameters on individual modes from some
 of the lightest stars in the Kepler field.

\end{itemize}

 \section{Stellar cycle effects}
 \label{sec:cyc}

Next, we consider the impact of stellar cycle effects on the p-mode
parameters of Sun-like oscillators.  From our experience of analyzing
helioseismic data we expect not only the mode frequencies, but also
the mode heights and linewidths, to show stellar-cycle variations.  We
begin by using scaling relations to predict the sizes of the
stellar-cycle changes for each member of our grid of stellar models.

 \subsection{Predicted stellar-cycle shifts}
 \label{sec:cycpred}

A commonly used indicator of surface activity on stars is the Ca~II
H\&K index. This index is usually expressed as $R'_{\rm HK}$, the
average fraction of the star's total luminosity that is emitted in the
H\&K line cores (having been corrected for the photospheric
component).  In order to give first-order estimates of the
stellar-cycle shifts in the p modes, we have made use of data on 22
main-sequence stars that have been observed by the Mount Wilson Ca~II
H\&K program (Radick et al. 1998; Saar \& Brandenburg 2002; see also
Baliunas et al. 1995) to have well-defined, and measurable, stellar
activity cycles in $R'_{\rm HK}$. These data, which come from Saar \&
Brandenburg, are the same data that were used in Chaplin et al. (2007)
and we refer the reader to that paper for more details. Here, we give
a brief summary of how we estimated mean p-mode frequency shifts from
these data.

First, we turned the $\Delta R'_{\rm HK}$ data (the cycle amplitudes
in $R'_{\rm HK}$) into first-order estimates of mean p-mode frequency
shifts by simply scaling against the 0.4-$\rm \mu Hz$ frequency shift
seen for the most prominent low-$l$ modes on the Sun. In doing so, we
assumed that the low-$l$ shifts scale approximately linearly with
$\Delta R'_{\rm HK}$. A linear fit of the estimated p-mode frequency
shifts versus the observed $R'_{\rm HK}$ then served as a look-up
curve to calculate a mean p-mode frequency shift for the most
prominent modes of each of the 31 models in this paper. There is some
uncertainty in our simple scaling due to the unknown inclination, $i$,
of the stars. We attempted to make some allowance for this by
weighting the fits, using the estimated uncertainties discussed in
Chaplin et al. (2007) as weights.

\begin{figure*}
 \centerline {\epsfxsize=8.0cm\epsfbox{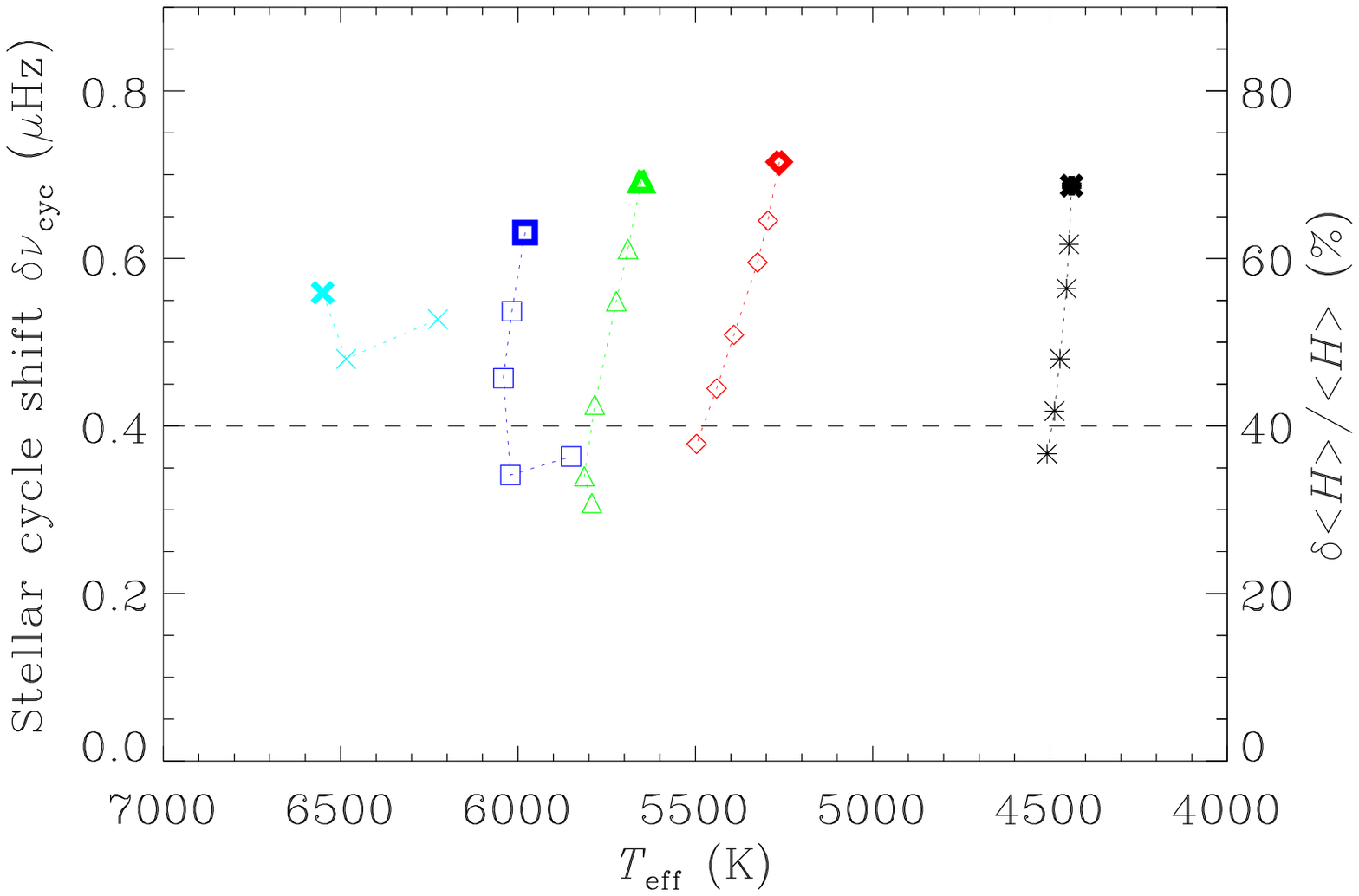}
              \epsfxsize=8.0cm\epsfbox{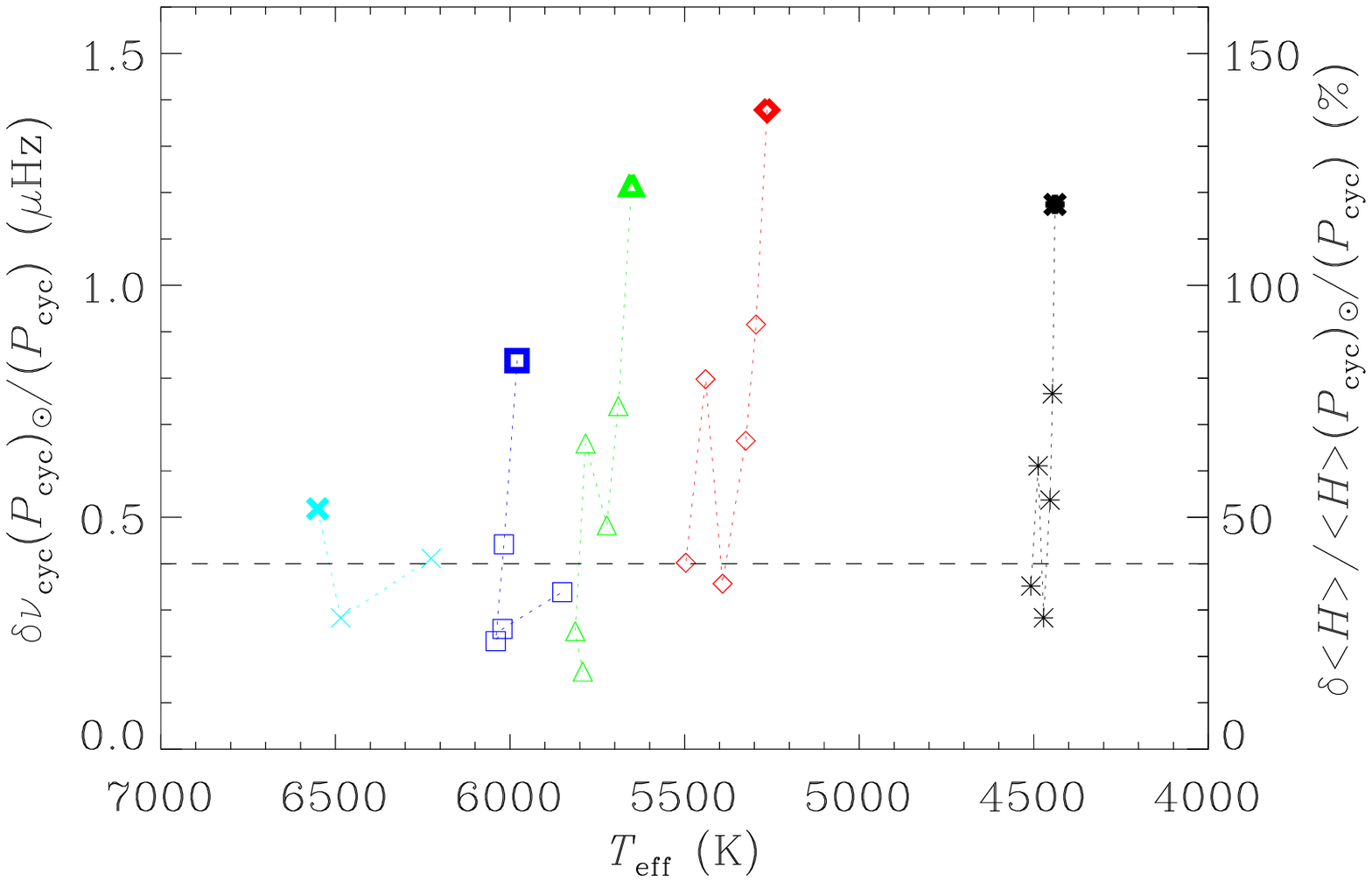}}
 \centerline {\epsfxsize=8.0cm\epsfbox{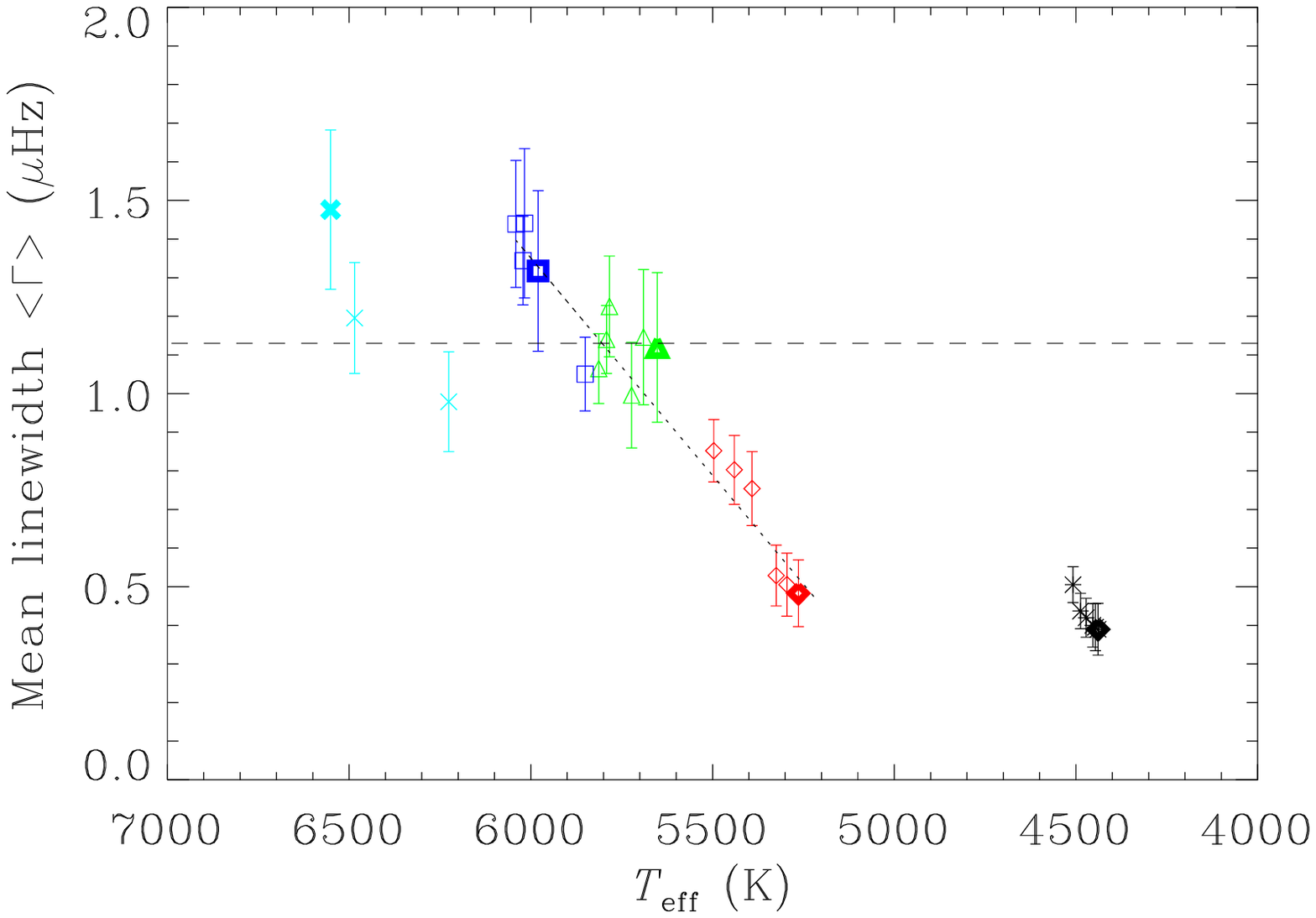}
              \epsfxsize=8.0cm\epsfbox{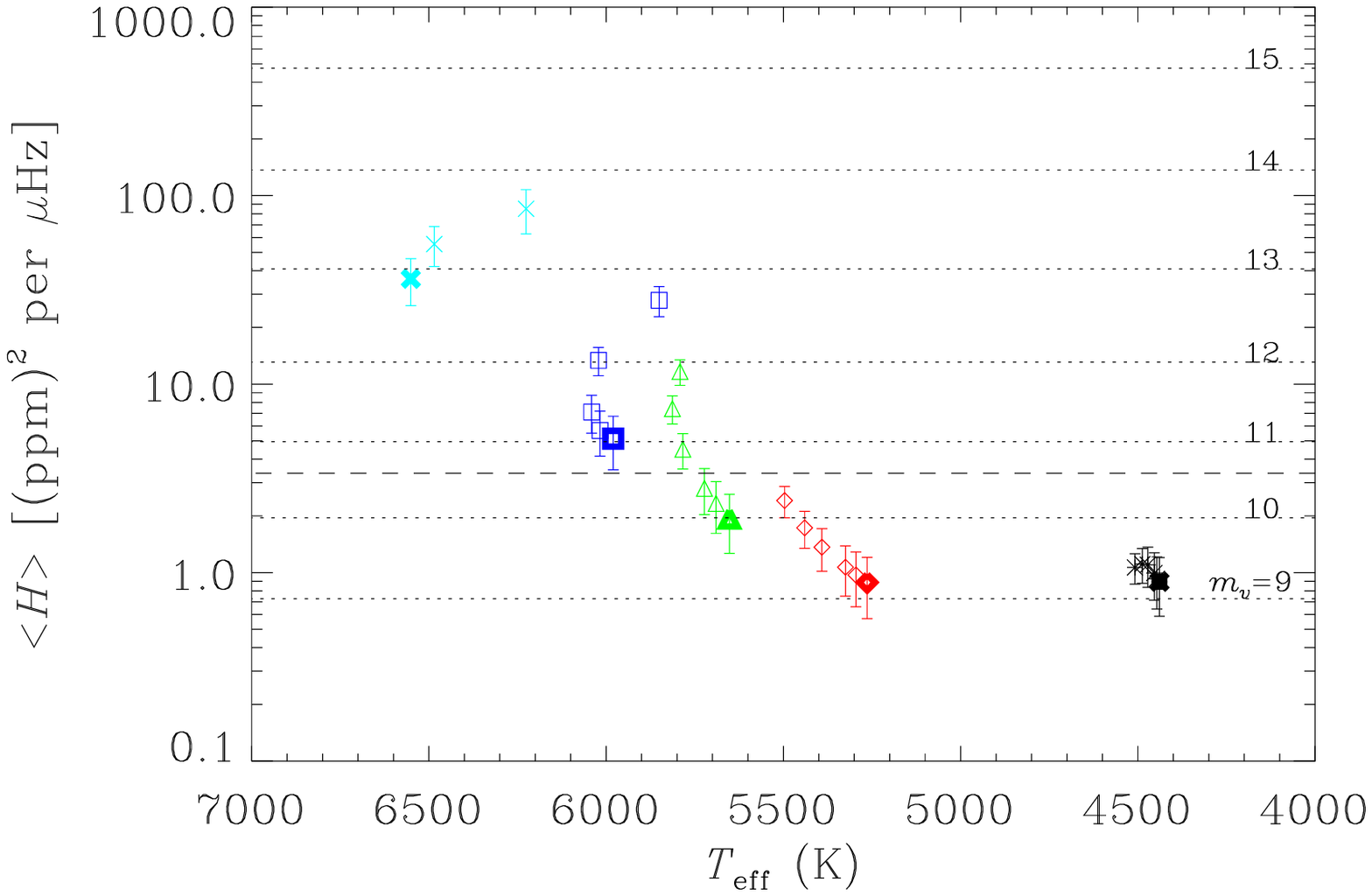}}

 \caption{Top left-hand panel: Predicted stellar-cycle frequency
 shifts. Top right-hand panel: Predicted shifts from the top left-hand
 panel, corrected for the stellar-cycle periods. Bottom panels:
 predicted and $\left< \Gamma \right>$ and $\left< H \right>$ from
 Fig.~\ref{fig:noi}, with error bars to show estimates of the lower
 and upper bounds on the parameters due to the predicted sizes of the
 stellar cycles. The dotted line in the bottom left-hand panel is the
 best-fitting straight line for fits to all models with $0.9 \le
 M/{\rm M_{\odot}} \le 1.1$. The dashed lines in each panel show
 averages for the five strongest solar radial modes. Data on the 1-Gyr
 models are rendered with bold symbols (ZAMS results are not shown;
 see text).}

 \label{fig:cyc}
\end{figure*}

In order to use the resulting look-up curve we first had to estimate
the $R'_{\rm HK}$ of each model.  We used scaling relations due to
Noyes (1983) and Noyes et al. (1984), which require $B-V$ and the
surface rate of rotation, to estimate the $R'_{\rm HK}$ of our model
stars. We discuss how we estimated the rotation rates in
Section~\ref{sec:rot} below. Estimated stellar-cycle frequency shifts
for the most prominent modes of the 31 models are plotted in the top
left-hand panel of Fig.~\ref{fig:cyc}. Because these predictions rely
on predictions of the surface rates of rotation, which are more
uncertain for the ZAMS models, the youngest data we show here are the
1-Gyr results (bold symbols). For each mass, one should pan
\emph{down} the symbols to go to older models. We therefore see that
the predicted amplitudes of the stellar cycles decrease monotonically
with age for all but the two most massive model sequences.

The right-hand axis of the top left-hand panel of Fig.~\ref{fig:cyc}
shows the scale for the predicted fractional changes in the average
heights, $\left< H \right>$, of the stellar models (those in the
average linewidths, $\left< \Gamma \right>$, are half the size). As
with the frequency shifts, we have scaled the shifts of the heights
and widths using the known solar values. The absolute fractional
change, from activity minimum to maximum, in the $\left< H \right>$ of
the low-$l$ solar p modes is about 40\,\%. That in $\left< \Gamma
\right>$ is about 20\,\% (Chaplin et al. 2000).

Chaplin et al. (2007) discussed at length the prospects for detecting
signatures of stellar activity cycles in the p-mode frequency
shifts. They concluded that provided modes covering several radial
orders are observed at reasonable S/N, it should be possible to
measure the parameter shifts given only a few multi-month segments of
data. The results here reinforce these conclusions: the top left-hand
panel of Fig.~\ref{fig:cyc} shows that most of the predicted
stellar-cycle frequency shifts have larger amplitudes than the Sun.

Prospects for detecting the stellar-cycle shifts from observations of
finite length depend not only on the amplitudes of the cycles, but
also on the cycle periods, $P_{\rm cyc}$. Take the case where the
observations span less than half the cycle period. Then for a given
cycle amplitude, the shorter the cycle period, the larger will be the
observed frequency shift. In order to take account of this dependence,
the top right-hand panel of Fig.~\ref{fig:cyc} plots the predicted
cycle shifts multiplied by the ratio $(P_{\rm cyc})_{\odot}/(P_{\rm
cyc})$. (Appendix~\ref{sec:pcyc} explains how we derived estimates of
$P_{\rm cyc}$.) The majority of these normalized shifts are again seen
to be larger than for the Sun.


The bottom panels of Fig.~\ref{fig:cyc} plot the average heights and
average linewidths, $\left< H \right>$ and $\left< \Gamma \right>$
respectively (Fig.~\ref{fig:noi}). However, this time we also use
error bars to show estimated lower and upper bounds on the parameters,
from our predictions of the stellar cycles.  The $\left< H \right>$ of
the younger models -- which in some cases have estimated cycle shifts
that are between one-and-a-half and two-times stronger than the Sun --
are predicted to vary by as much as $\approx 70$\,\%. When we make
detailed comparisons of predictions of $\left< H \right>$ -- from the
analytical pulsation calculations, or from numerical simulations --
with observations of $\left< H \right>$, we therefore need to be sure
we know whether stars show significant variations in activity, and if
they do, at what point in any cycles the observations have been
made. If information on the stellar cycles is not known a priori, our
predictions imply that we cannot ensure the accuracy of any
comparisons to better than (on average) about 50\,\% in $\left< H
\right>$.

Finally in this section, we note that the predicted widths in the
bottom left-hand panel of Fig.~\ref{fig:cyc} show an approximate
linear dependence on the effective temperature, $T_{\rm eff}$, in the
range from $0.9\,\rm M_{\odot}$ to $1.1\,\rm M_{\odot}$. The dotted
line shows the best-fitting straight line for all models in this
range. We used the error bars from the stellar-cycle predictions as
weights for the straight-line fit, which is described by:
 \begin{equation}
 \left< \Gamma \right> =-(5.4 \pm 0.8) + (11.3 \pm 1.5) 
 \times 10^{-4}\,T_{\rm eff}\,\,\rm \mu Hz.
 \label{eq:widpred}
 \end{equation}
Equation~\ref{eq:widpred} may be used to make predictions of the
linewidths of the most prominent p modes of models having $0.9 \la
M/{\rm M_{\odot}} \la 1.1$.

 \subsection{Distortion of mode peaks by stellar cycles}
 \label{sec:distort}

Large stellar-cycle frequency shifts can create potential problems for
peak-bagging. The problems arise when the frequency power spectrum to
be analyzed is made from a timeseries within which significant
variation of the frequencies is present. The resonant peaks in the
frequency power spectrum may then be distorted because of the
frequency variations. The distortion means the peaks no longer have
the Lorentzian-like shapes, on which the peak-bagging fitting models
are based. These issues are discussed at length in Chaplin et
al. (2008), where results on the expected distortions are presented
for the solar case. While the distortions have little impact on
estimates of the mode frequencies, they can bias estimation of the
heights and widths, because the Lorentzian fitting models are no
longer a good representation of the underlying peaks.

\begin{figure*}
 \centerline{\epsfxsize=8.0cm\epsfbox{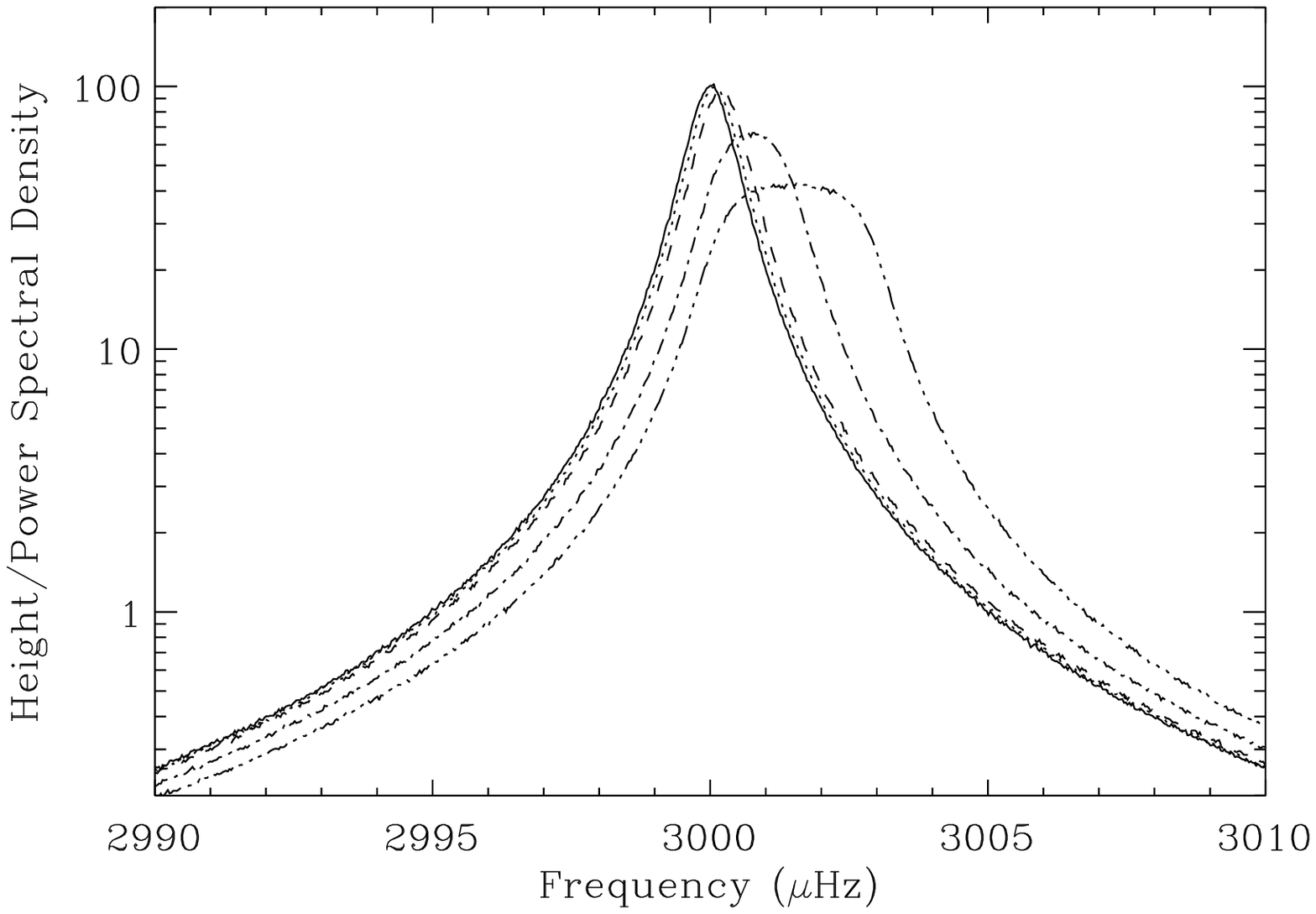}}
 \centerline {\epsfxsize=8.0cm\epsfbox{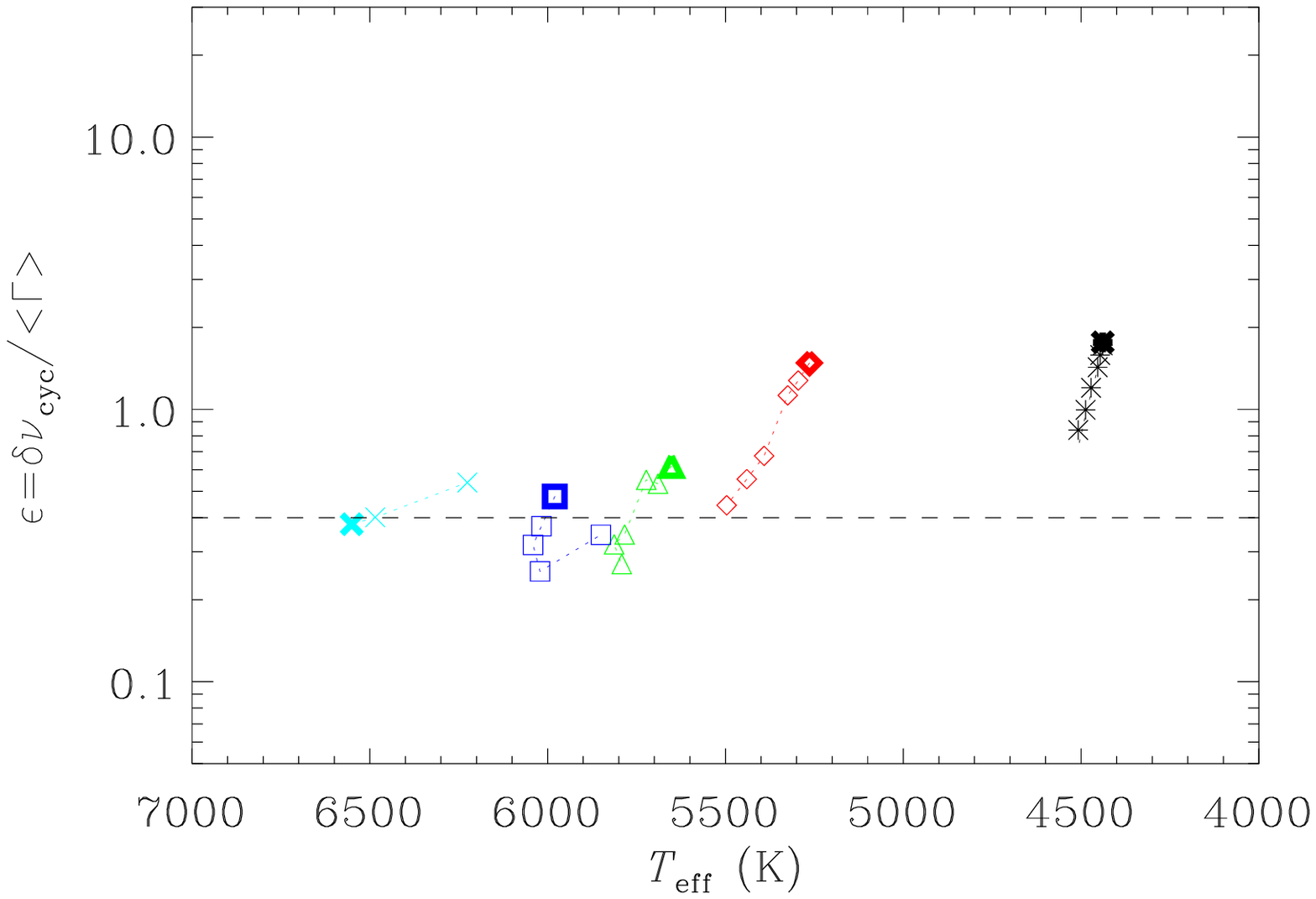}
              \epsfxsize=8.0cm\epsfbox{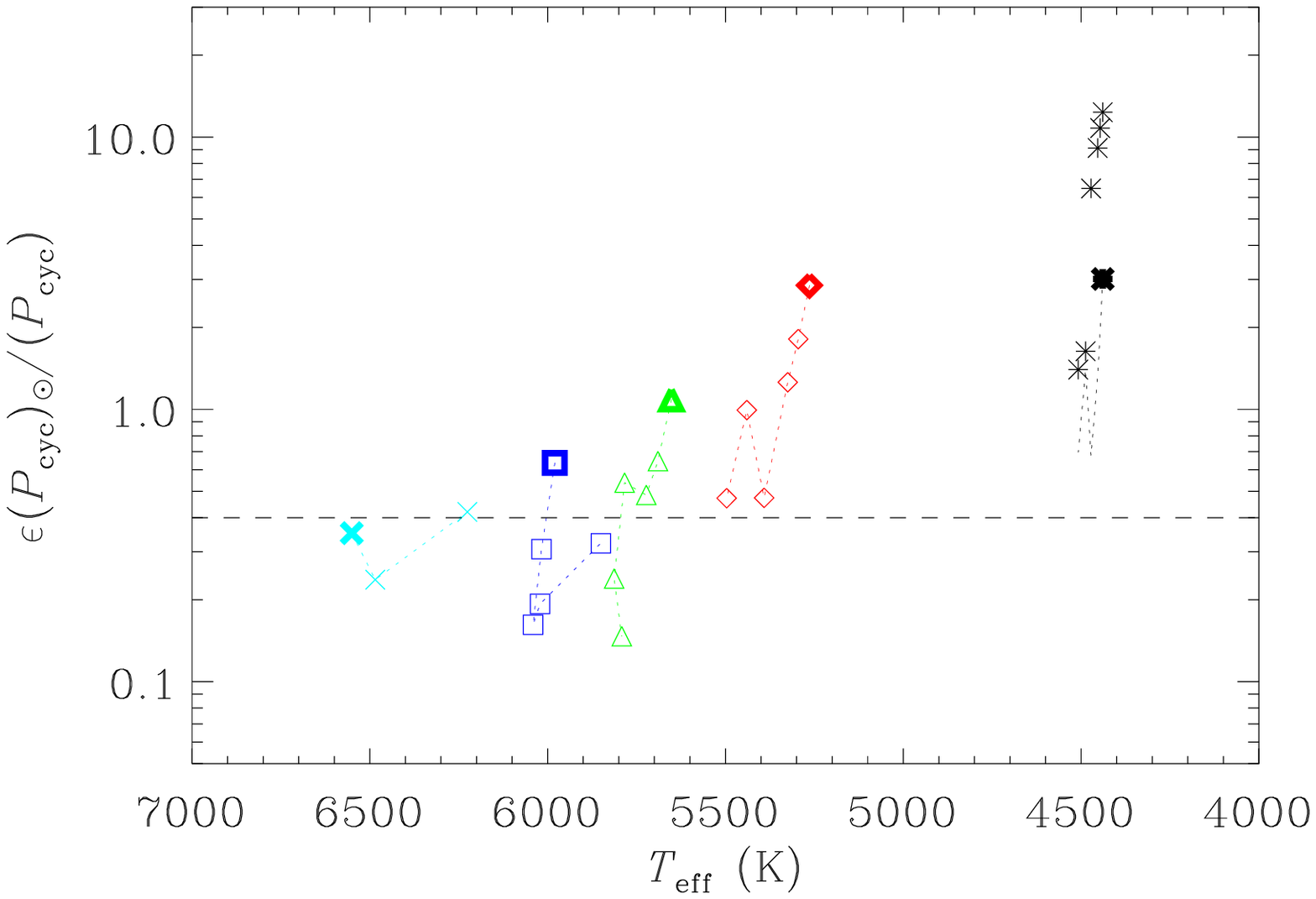}}

 \caption{Top panel: Peak profiles expected for a single mode of width
 $\Gamma=1\,\rm \mu Hz$, in the frequency power spectrum of a time
 series within which the frequency was varied in a linear manner by
 total amount $\delta\nu_{\rm cyc}$ (figure from Chaplin et al. 2008).
 The distortion of the peak profile depends on the ratio
 $\epsilon=\delta\nu_{\rm cyc}/\left< \Gamma \right>$. Various
 linestyles show peak profiles for: no shift (solid line);
 $\epsilon=0.15$ (dotted line); 0.4 (dashed line); 1.5 (dot-dashed
 line); and 3.0 (dot-dot-dot-dashed line). Bottom left-hand panel:
 Ratios, $\epsilon$, of the predicted stellar-cycle frequency shifts
 and the average linewidths, $\left< \Gamma \right>$. Bottom
 right-hand panel: The ratios, $\epsilon$, corrected for the
 stellar-cycle periods. The dashed lines in the bottom panels show
 averages for the five strongest solar radial modes. Data on the 1-Gyr
 models are rendered with bold symbols (ZAMS results are not shown).}

 \label{fig:cycsh}
\end{figure*}

It turns out that the key parameter for determining the extent of any
distortion is the ratio of the frequency shift, $\delta\nu_{\rm cyc}$,
in the timeseries to the peak linewidth. We call this ratio
$\epsilon$. Some examples of distorted peaks are shown in the top
panel of Fig.~\ref{fig:cycsh}, for values of $\epsilon=0.0$, 0.15,
0.40, 1.50 and 3.00.  Note that $\epsilon \approx 0.4$ for the most
prominent low-$l$ solar p modes (for which $\delta\nu_{\rm cyc}
\approx 0.4\,\rm \mu Hz$ and $\left< \Gamma \right> \approx 1\,\rm \mu
Hz$).

Aside from the change to the shape of the profiles, there are also
clearly implications for mode detectability. The profiles
characterized by $\epsilon=1.5$ and 3 have maximum heights that are,
respectively, only $\approx 60$\,\% and 40\,\% of the height of the
undistorted profile.

The bottom left-hand panel of Fig.~\ref{fig:cycsh} plots the estimated
$\epsilon=\delta\nu_{\rm cyc}/\left< \Gamma \right>$ of our model
stars. The dashed line shows the solar value. We see that the
prediction implies cooler stars are more likely to show larger peak
distortions. That said, we need to think a little harder about the
prediction. When we compare one star with another, the estimated
$\delta\nu_{\rm cyc}$, which are used to compute $\epsilon$, do not
tell the whole story since the frequency shifts observed in a
timeseries of given length will also depend (as noted previously) on
the cycle periods, $P_{\rm cyc}$, and those periods will vary from one
star to another.  In order to take account of this dependence on
$P_{\rm cyc}$, the bottom right-hand panel of Fig.~\ref{fig:cycsh}
plots $\epsilon$ multiplied by the ratio $(P_{\rm
cyc})_{\odot}/(P_{\rm cyc})$ (i.e., the shorter the cycle period, the
more severe the distortion effect for a given cycle amplitude). While
some of the detail of the plot is altered when we allow for the cycle
length, the overall trend of the results is maintained: we should
watch out for distortion effects in stars cooler than the
Sun. Moreover, as noted above, these distortions may have an impact on
the detectability of peaks in the frequency power spectrum.


 \section{Rotation}
 \label{sec:rot}

We have used empirical relations in the literature to calculate
surface rotation periods, $P_{\rm rot}$, for the models, which we have
then turned into equivalent rotational frequency splittings, $\delta
\nu_{\rm rot} = 1/P_{\rm rot}$. We initially compared two empirical
relations, due to Aigrain et al. (2004) and Cardini \& Cassatella
(2007), respectively. Note that in what follows we have not plotted
the results for the ZAMS models, since the empirical relations are
more uncertain there. Plotted data therefore span the age range from 1
to $\sim 9\,\rm Gyr$.

\begin{figure*}
 \centerline {\epsfxsize=8.0cm\epsfbox{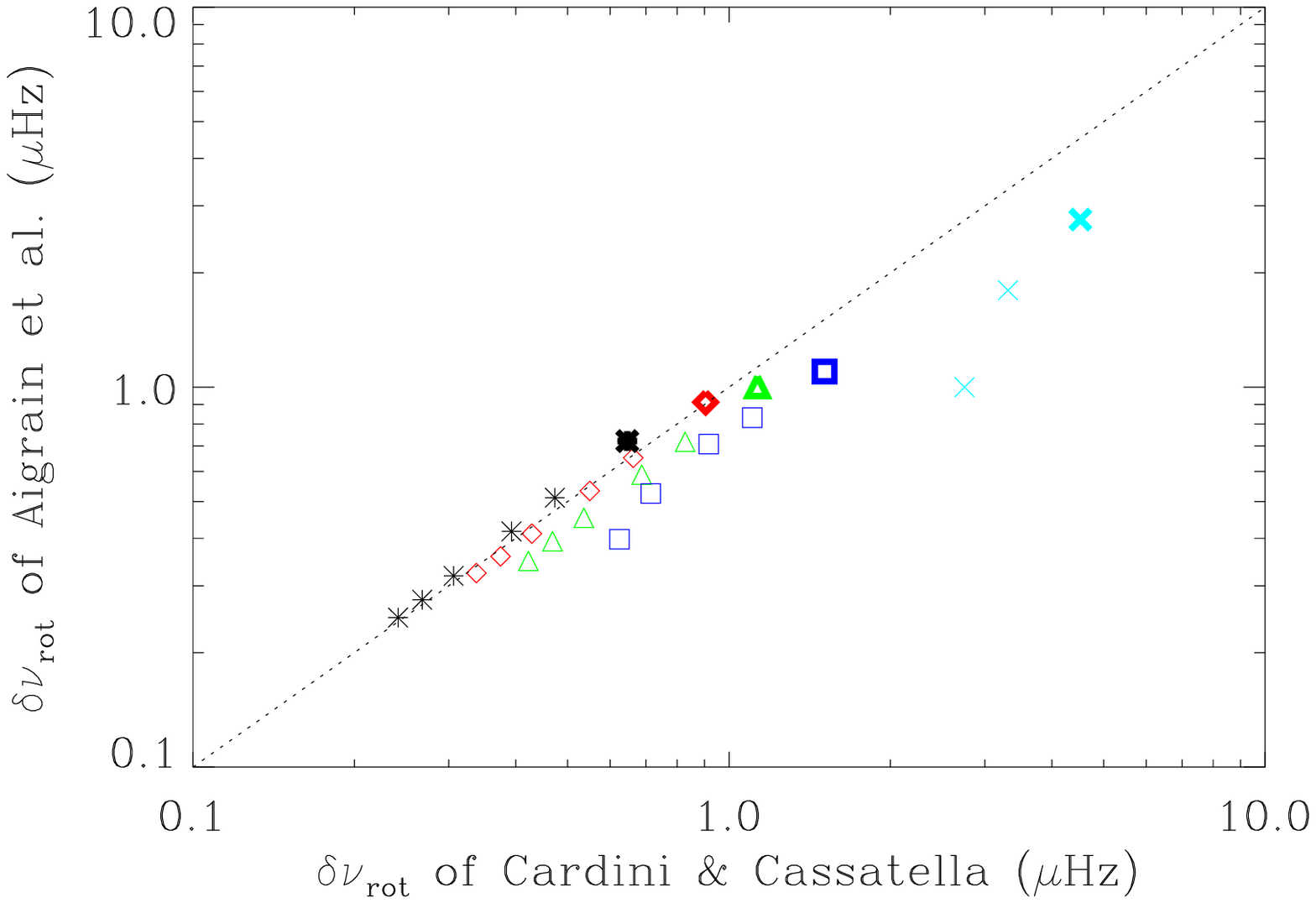}
              \epsfxsize=8.0cm\epsfbox{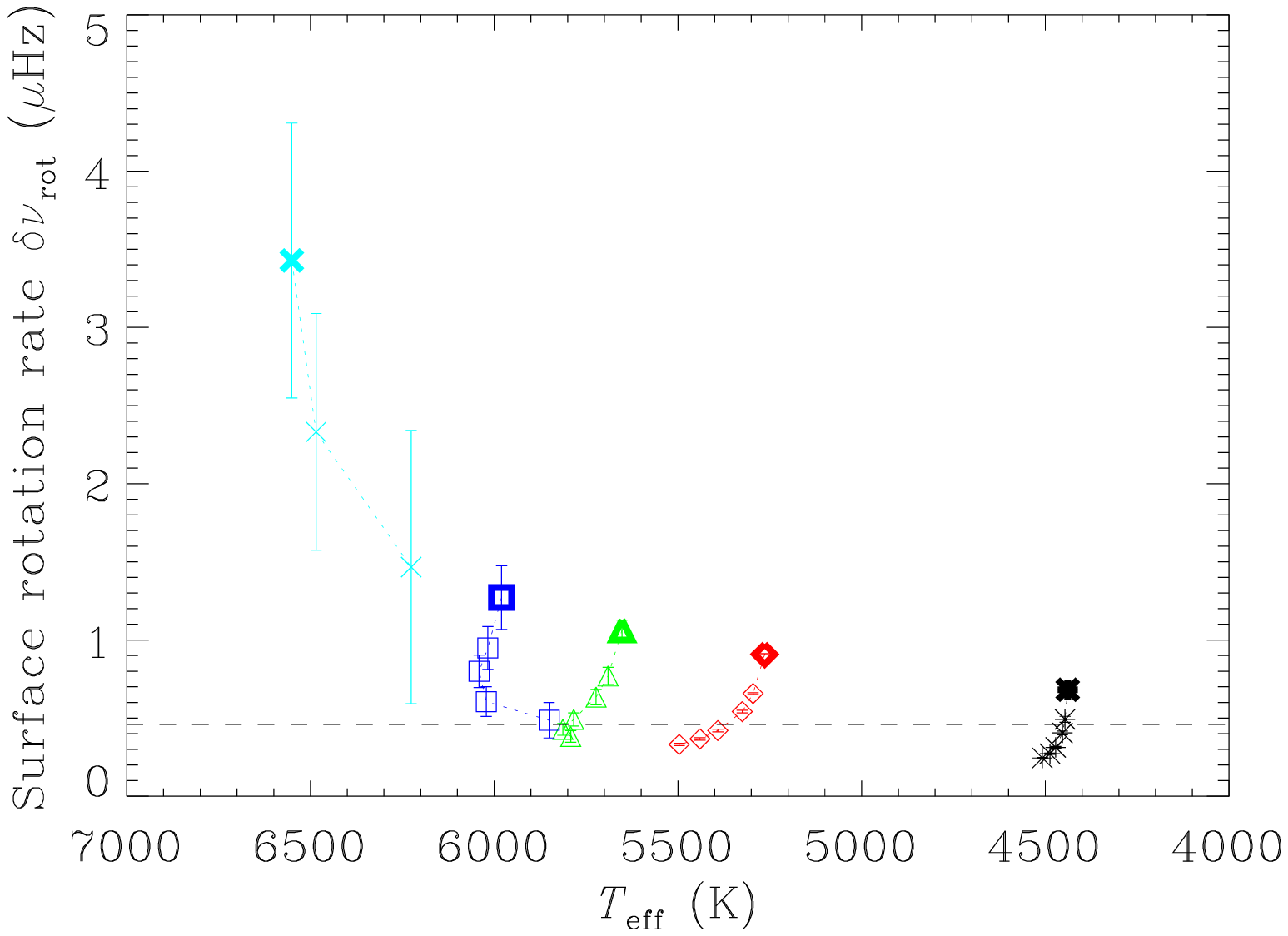}}
 \centerline {\epsfxsize=8.0cm\epsfbox{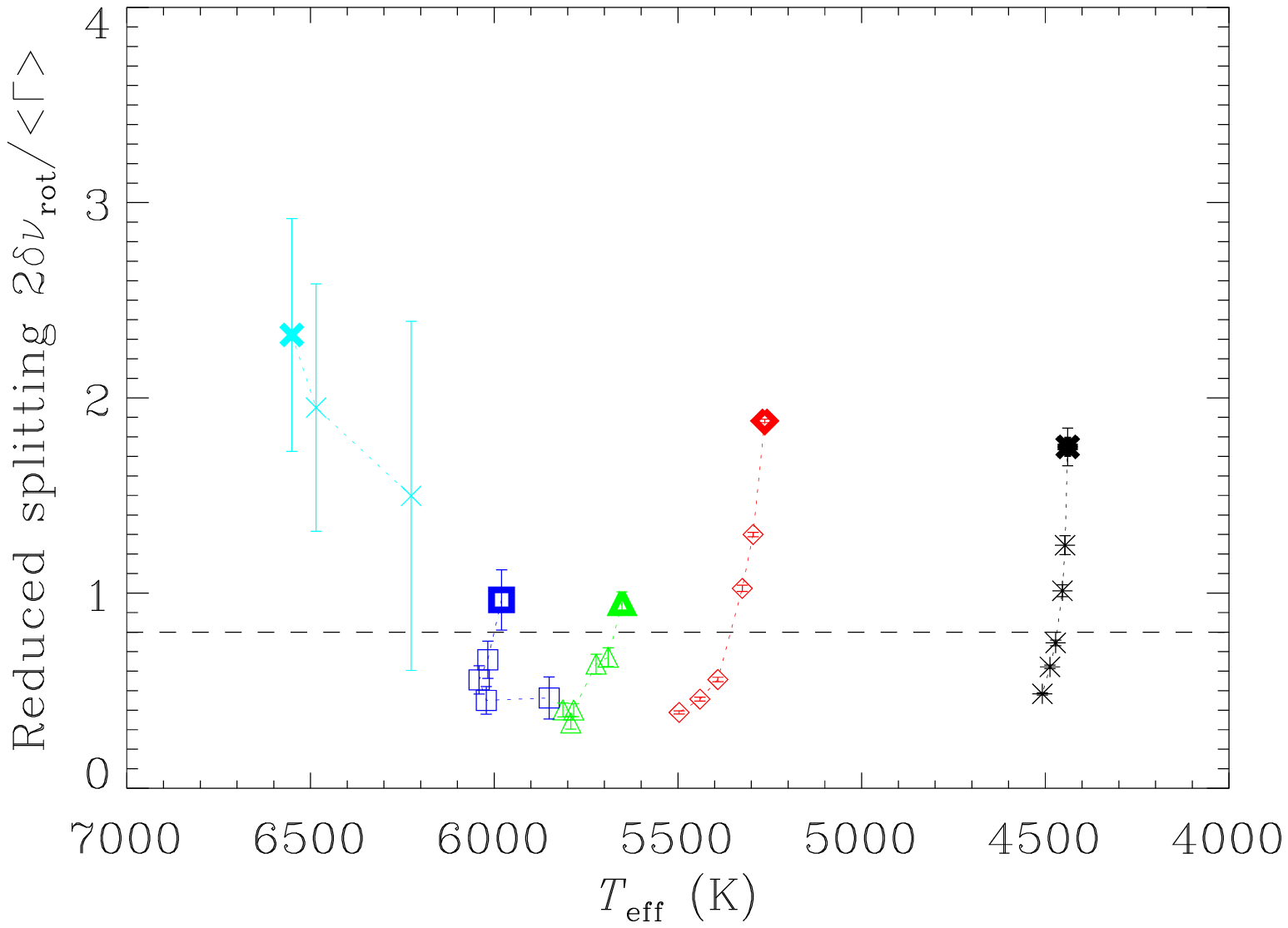}}

 \caption{Top left-hand panel: Comparison of predicted surface
 rotation rates, here shown as equivalent rotational frequency
 splittings, from Aigrain et al. (2004) and Cardini \& Cassatella
 (2007). Top right-hand panel: Frequency splittings, $\delta \nu_{\rm
 rot}$, for predicted surface rates of rotation. Bottom panel: reduced
 splittings, $2\delta \nu_{\rm rot}/\left< \Gamma \right>$. The dashed
 lines in top-right and bottom panels show averages for the five
 strongest solar radial modes. Data on the 1-Gyr models are rendered
 with bold symbols (ZAMS results are not shown).}

 \label{fig:rot}
\end{figure*}

The Aigrain et al. relation makes explicit use of the $B-V$ colour and
age to determine $P_{\rm rot}$; while the Cardini \& Cassatella
relation makes explicit use of the mass and age. The left-hand panel
of Fig.~\ref{fig:rot} plots estimates of the $\delta \nu_{\rm rot}$ of
the models against one another. Reasonable agreement is seen, apart
from for the highest-mass $1.3\,\rm M_{\odot}$ models, where the
Cardini \& Cassatella estimates are higher than their Aigrain et
al. counterparts.


In what follows we have averaged the estimated rotation periods of the
two methods, and used differences between the estimates as a measure
of uncertainty for the predictions. The top right-hand panel of
Fig.~\ref{fig:rot} shows averaged estimates of $\delta \nu_{\rm rot}$
as a function of effective temperature. The error bars indicate the
sizes of the differences between the predictions given by the two
methods. The dashed line shows the splitting for the Sun.  The data in
this panel of course contain the expected trends: at fixed age, cooler
stars have slower surface rates of rotation; while at fixed $T_{\rm
eff}$, the older the star, the slower is the rate of rotation (i.e.,
for each mass, one should pan \emph{down} the symbols to go to older
models).

It is at this point that our predictions become more speculative. This
is because the data in Fig.~\ref{fig:rot} are estimated rotational
splittings of the \emph{surface} layers, not the internal global
averages that the low-$l$ p-mode splittings will measure. To progress
we can choose to assume that, like the Sun, stars possess mean
internal rotation rates that are comparable to the surface rates, an
assumption which is very likely even more questionable for the younger
stars. We may then use the splittings in the top right-hand panel of
Fig.~\ref{fig:rot} as a rough guide to typical splittings we might
encounter in the frequency power spectra. It would seem reasonable to
at least treat these splittings as lower-limit estimates on the
internal global splittings, assuming internal rates are unlikely to be
lower than the surface rates.

Next, we define a reduced splitting, $2\delta \nu_{\rm rot}/\left<
\Gamma \right>$, which is the ratio of twice the rotational frequency
splitting to the mean linewidth. As already noted in
Section~\ref{sec:kepler}, this ratio determines how easy it is to
resolve adjacent components in the non-radial modes and is therefore
an important indicator of when things get more difficult for the
peak-bagging codes. The smaller is the ratio, the more blended
adjacent components in the multiplets will be, and the more difficult
extraction of the frequency splitting and the angle of inclination
will become.

The reduced splittings of the models are plotted in the bottom panel
of Fig.~\ref{fig:rot}. We take three important points from the
plot. First, the tendency is for the reduced splittings to be larger,
and hence potential difficulties for the peak-bagging to be less
severe, at low and high $T_{\rm eff}$ (in particular for the younger
models). Second, a good fraction of the models in the central part of
the plot actually have reduced splittings that are \emph{smaller} than
for the Sun, which is potentially problematic. And third, the results
show that any problems will get worse as stars age on the main
sequence.

\section{Summary}
\label{sec:sum}

The main points of the paper may be summarized as follows:

\begin{enumerate}

 \item We used analytical pulsation computations of the excitation and
 damping properties of p modes, together with estimated shot noise
 levels, to make predictions of p-mode detectability in future
 observations of main sequence stars by NASA's Kepler mission. The
 computations tested models of stars having masses in the range from
 $0.7\,\rm M_{\odot}$ to $1.3\,\rm M_{\odot}$.

 The predictions (see Fig~\ref{fig:noi}) suggest it should be possible
 to detect individual modes at reasonable S/N when the target stars
 have similar, or higher, intrinsic brightness than the Sun. Stars that
 are intrinsically fainter than the Sun will be more of a challenge
 for the analysis. Stars as light as $0.7\,\rm M_{\odot}$ -- which
 have luminosities $\approx 0.15\,\rm L_{\odot}$ -- may have p-mode
 powers so small as to severely limit robust parameter extraction
 on individual modes.
 
  \item Our predictions of variability due to the stellar activity
 cycles -- in the age range of roughly 1 to 9\,Gyr -- suggest
 variability of p-mode frequencies, heights and linewidths can be up
 to one-and-a-half to two-times as strong as the Sun (see
 Fig.~\ref{fig:cyc}).  While the predictions show little change in the
 size of the variability with stellar mass (or $T_{\rm eff}$), at
 fixed mass the variability decreases with increasing age. Our
 predictions emphasize the need to take into account any cyclic
 variability of the mode powers and linewidths, when theoretical
 predictions of those powers and linewidths are compared with the
 observations.

 \item Stellar activity cycles may in some stars distort the shapes of
 mode peaks in the frequency power spectrum (see
 Fig.~\ref{fig:cycsh}). The distortion depends on the ratio of the
 stellar-cycle frequency shift in the timeseries to the mode
 linewidth. Our stellar-cycle predictions suggest stars cooler than
 the Sun are most susceptible to the effect. The distortion has two
 important consequences. First, if no account is made of the
 distortion it can lead to poor estimates of the power and linewidth
 parameters. Second, it can lead to significant reduction of the
 heights of mode peaks (peaks are `squashed' by the distortion),
 making detection of modes more difficult in the frequency power
 spectrum.

 \item We also looked at the ability to resolve individual p-mode
 components in the non-radial mode multiplets, the success of which
 depends on the relative importance of rotation and mode damping (see
 Fig.~\ref{fig:rot}). This is an important aspect of any analysis on
 individual modes, in that it in principle permits estimation of the
 angle of inclination offered by the star, and also robust extraction
 of rotational frequency splittings, and possibly also any asymmetry
 of those splittings due to stellar activity. Our predictions suggest
 resolution problems may be less severe than the solar case in stars a
 few-hundred-Kelvins hotter than ($M \ga 1.3\,\rm M_{\odot}$), or
 cooler than ($M \la 0.9\,\rm M_{\odot}$), the Sun.  Resolution
 problems may be most severe for stars having similar $T_{\rm eff}$
 (and mass) to the Sun. There is also a marked tendency for the
 problems to get worse as stars age on the main sequence. It is worth
 adding how difficult it has proven to get reliable estimates of the
 splittings from Sun-as-a-star data (e.g., see Chaplin et al. 2006,
 and references therein).

\end{enumerate}

\begin{acknowledgements}

This work came out of the
asteroFLAG\footnote{http://www.issibern.ch/teams/Astflag} project, in
particular its participation in preparations for Kepler.  We
acknowledge the International Space Science Institute (ISSI), which
provides support for asteroFLAG. This work was also supported by the
European Helio- and Asteroseismology Network
(HELAS)\footnote{http://www.helas-eu.org}, a major international
collaboration funded by the European Commission's Sixth Framework
Programme. We would like to thank H.~Kjeldsen for useful comments and
also (with W.~J.~Borucki) for providing the data on estimated noise
levels for Kepler. We also thank the referee for helpful comments on
the draft. WJC, GH, RN and TT also acknowledge the support of STFC.

\end{acknowledgements}

\appendix

\section{Dependence of $H$ on length of dataset}
\label{sec:hall}

Equation~\ref{eq:height} in Section~\ref{sec:sn} strictly only gives
the correct value for $H$ as the length of the observation, $T$, tends
to infinity. The proper description of $H$, which also covers the
regime where mode peaks are unresolved, is actually given by (Fletcher
et al. 2006):
 \begin{equation}
 H(T) = \frac{2A^2T}{\pi \Gamma T +2}.
 \label{eq:hfull}
 \end{equation}
When $\pi T \Gamma << 2$ (i.e., when $T << 2\tau$) the mode is not
resolved, and power is largely confined in one bin of the frequency
power spectrum (so that $H \sim A^2$). On the other hand, when $\pi T
\Gamma >> 2$ (i.e., when $T >> 2\tau$), the mode is well resolved, and
the description approaches Equation~\ref{eq:height}.

\begin{figure}
 \centerline {\epsfxsize=8.0cm\epsfbox{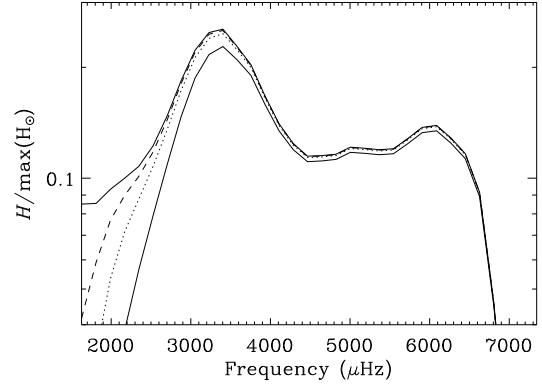}}

 \caption{Predictions of observed heights of modes for $M=0.9\,\rm
 M_{\odot}$ model of age 5\,Gyr. The curves show estimates of heights
 for different lengths of observation, $T$: thin solid line for $T=93\,\rm
 days$; dotted for $T=1\,\rm yr$; and dashed for $T=3.5\,\rm yr$; and
 thick solid line for $T=\infty$, i.e., $H_{\infty}$.}

 \label{fig:kepht}
\end{figure}

What happens in the intermediate regime, where $\pi T \Gamma$ is
neither much greater, or much smaller, in size than 2? The form of
Equation~\ref{eq:hfull} indicates there is a gradual transition
between the $H$ for the unresolved, and fully resolved regimes. To see
more clearly how $H$ is affected, let us re-tag the height given by
Equation~\ref{eq:height} as $H_{\infty}$, i.e.,
 \[
 H_{\infty} = \frac{2A^2}{\pi\Gamma}.
 \]
Then, the more general $H(T)$ from Equation~\ref{eq:hfull} may be
written as:
 \begin{equation}
 H(T) = \frac{H_{\infty}}{\left[ 1 + \left(2 / \pi \Gamma T \right) \right]}.
 \label{eq:hrel}
 \end{equation}
Equation~\ref{eq:hrel} therefore shows how the observed height
compares to $H_{\infty}$ as $T$ is varied. Fig.~\ref{fig:kepht} shows
representative examples for the $M=0.9\,\rm M_{\odot}$ model of age
5.25\,Gyr. The curves show estimates of heights for different lengths of
observation, these being $T=93\,\rm days$ (thin solid line), $T=1\,\rm
yr$ (dotted line) and $T=3.5\,\rm yr$ (dashed line) respectively. The
dark solid line shows the heights for $T=\infty$ (i.e., $H_{\infty}$).

At the lowest frequencies the modes are not well resolved, and the
lifetimes $\tau$ are in some cases comparable to $T$ (meaning $\pi T
\Gamma \sim 2$). The observed heights of these modes are therefore
affected significantly by the increase in $T$ from 93\,days to
1\,yr. What is interesting is that the heights at maximum power also
show effects from the change in $T$. Here, lifetimes are of order
$\approx 5\,\rm days$. This means that even for the shortest
observation time ones' initial reaction would be that the modes seem
to be reasonably well resolved, i.e., we have $T \approx 20 \tau$.
The observed heights for $T$ from 93\,days are nevertheless about
10\,\% smaller than the $H_{\infty}$. Differences at the longer
observation lengths are rather more modest. It is worth adding that
these differences are smaller than the typical uncertainties implied
by the precision achievable in the parameters (e.g., see the formulae
in Toutain \& Appourchaux 1994).


As far as the predictions in the main part of this paper are concerned
-- which are for modes around maximum power in each frequency power
spectrum, and which use the expression for $H_{\infty}$ -- differences
of 10\,\% or so are not a major issue. However, the discussion here
shows that detailed, frequency-dependent comparisons would be better
made with Equations~\ref{eq:hfull} and~\ref{eq:hrel}, in particular
for estimation of the visibilities of the more lightly damped modes.

 \section{Scaling relation to predict stellar-cycle period, $P_{\rm cyc}$}
 \label{sec:pcyc}

The scaling relations used to make data for this paper are all in the
literature, apart from one: the scaling relation we used to predict
the stellar-cycle periods, $P_{\rm cyc}$. In this section we explain
how this scaling relation was derived.

We again made use of activity and stellar-cycle data on 22
main-sequence stars that have been observed by the Mount Wilson Ca~II
H\&K program to have well-defined, and measurable, stellar activity
cycles (Saar \& Brandenburg 2002; see also Section~\ref{sec:cycpred}
and Chaplin et al. 2007).

\begin{figure*}
 \centerline{\epsfxsize=8.0cm\epsfbox{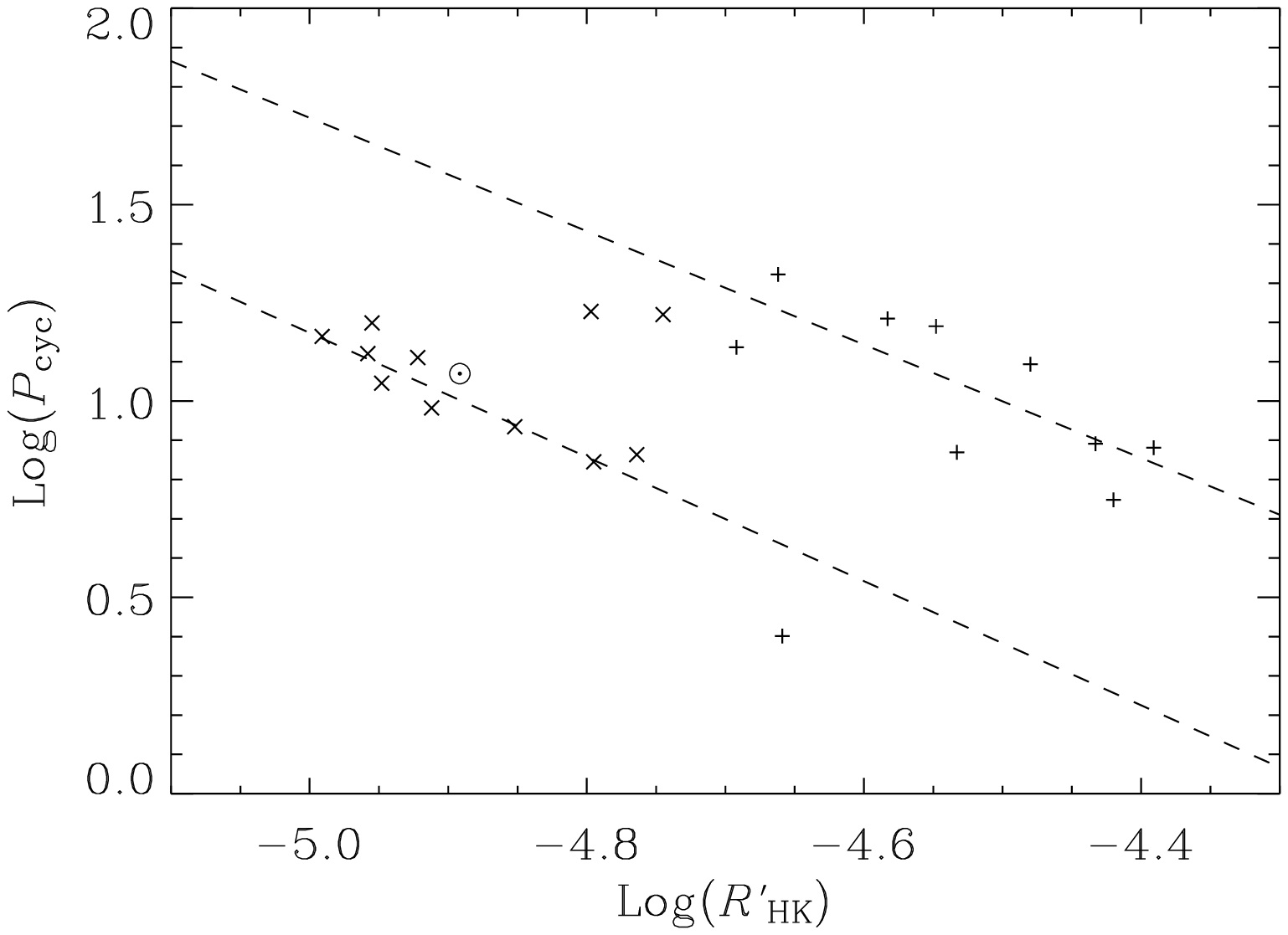}
             \epsfxsize=8.0cm\epsfbox{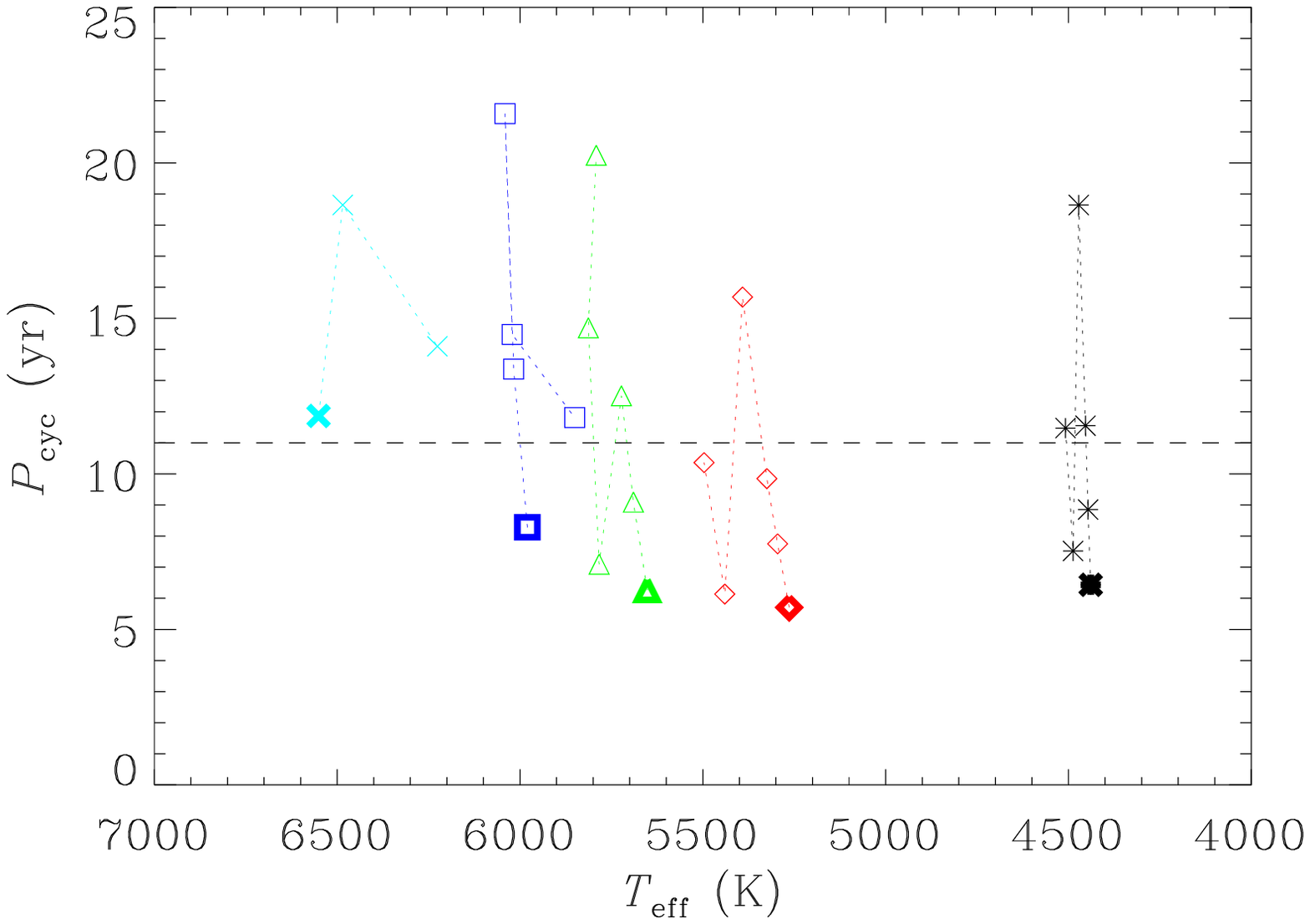}}

 \caption{Left-hand panel: The logarithm of the measured stellar-cycle
 period, $\log(P_{\rm cyc})$ against $\log(R'_{\rm HK})$, for data on
 22 stars in Saar \& Brandenburg (2002) [see also Chaplin et
 al. (2007)]. Two branches are seen to emerge from the plot. The
 dashed lines are best-fitting power-law fits to the two branches. The
 Sun is plotted with its usual symbol. Stars with $\log{R'_{\rm HK}} >
 -4.75$ are rendered as pluses; while stars with $\log{R'_{\rm HK}}
 \le -4.75$ are shown as crosses. The Sun falls in the second,
 less-active cohort. Right-hand panel: Estimated $P_{\rm cyc}$ for the
 31 stellar models in this paper, as derived by use of
 Equation~\ref{eq:plaw2}, which describes the power-law fits plotted
 in the left-hand panel. The dashed line marks the 11-year cycle
 period of the the Sun.}

 \label{fig:ppcyc}
 \end{figure*}

The left-hand panel of Fig.~\ref{fig:ppcyc} plots the logarithm of the
measured cycle periods of the 22 stars (in years), against the
logarithm of their measured activity indices, $R'_{\rm HK}$ (see
Section~\ref{sec:cycpred}). As in Saar \& Brandenburg, the stars have
been divided into two cohorts: `active' stars, plotted as pluses, for
which $\log{R'_{\rm HK}} > -4.75$; and `inactive' stars, plotted as
crosses, for which $\log{R'_{\rm HK}} \le -4.75$. The Sun, which is
plotted with its usual symbol, falls in the less-active cohort.


Two `branches' are seen to emerge in the left-hand panel of
Fig.~\ref{fig:ppcyc}. One branch contains mainly the `active' stars;
while the other branch contains mainly the `inactive' stars. This
division of activity data into two branches is also seen in other
combinations of parameter choices (e.g., see Saar \& Brandenburg).

The dashed lines are the best-fitting power laws for each branch. Both
best-fitting power-law indices are significant at $\approx
4\sigma$. The fits are described by:
 \begin{equation}
 \mbox{$\log(P_{\rm cyc})=$} \left\{
 \begin{array}{ll}
 \mbox{$-6.7-1.6\log(R'_{\rm HK})$}& \mbox{for $\log(R'_{\rm HK}) < -4.75$}\\
 \mbox{$-5.5-1.4\log(R'_{\rm HK})$}& \mbox{for $\log(R'_{\rm HK}) \ge -4.75$}\\
 \end{array} \right.
 \label{eq:plaw2}
 \end{equation}
Equation~\ref{eq:plaw2} was used to derive estimated cycle periods for
the 31 models in this paper (the $R'_{\rm HK}$ having already been
determined; see Section~\ref{sec:cycpred}). The estimates are plotted
in the right-hand panel of Fig.~\ref{fig:ppcyc} (same symbol and
colour scheme as in the rest of the paper). The overall tendency is
for the cycle period to lengthen as stars age. The irregular
appearance of the evolutionary sequences is due to the fact that some
models which are adjacent in age have values of $\log(R'_{\rm HK})$
that lie on either side of the $\log(R'_{\rm HK}) = -4.75$ boundary.

\end{document}